\documentclass[twocolumn]{aastex631}

\newcommand{\vhelio}{$v_{\rm helio}$}
\newcommand{\kms}{km s$^{-1}$}
\newcommand{\vHIlos}{$v_{\rm HI,los}$}
\newcommand{\veloffset}{$v_{\rm offset}$}
\newcommand{\psb}{$p_{\rm M32, SB}$}
\newcommand{\reff}{$r_{\rm eff}$}
\newcommand{\sigca}{$\Sigma$Ca}
\newcommand{\signa}{$\Sigma$Na}
\newcommand{\fehcat}{[Fe/H]$_{\rm CaT}$}
\newcommand{\fehphot}{[Fe/H]$_{\rm phot}$}
\newcommand{\sigmafehphot}{$\sigma_{\rm [Fe/H]}$}
\newcommand{\pvel}{$p_{\rm M32, vel}$}

\newcommand{\Nnomem}{63}
\newcommand{\ntargets}{2687}
\newcommand{\nnophastgoodposrv}{544}
\newcommand{\ngiants}{2505}
\newcommand{\fmwcontamsample}{4.8\%}
\newcommand{\vtotmed}{4.1}
\newcommand{\snmed}{5.3}
\newcommand{\nrgb}{1486}
\newcommand{\nagb}{539}
\newcommand{\nmw}{119}
\newcommand{\vlospredmed}{$-380.8$}

\newcommand{\rammin}{0.9}
\newcommand{\rammax}{15.5}

\newcommand{\rmajeff}{23}
\newcommand{\rmineff}{30}

\newcommand{\percentbulge}{2.5\%}
\newcommand{\percentdisk}{92.6\%}
\newcommand{\percenthalo}{4.9\%}
\newcommand{\percenthaloliketot}{7.4\%}

\newcommand{\npne}{28}

\newcommand{\fmwcontam}{0.2\%}
\newcommand{\medrederr}{0.006}
\newcommand{\medblueerr}{0.03}

\newcommand{\rprojmin}{3.4}
\newcommand{\rprojmax}{7.7}
\newcommand{\fvhalonoprior}{5\%}
\newcommand{\sigmadisknofhprior}{110}
\newcommand{\frotnofhprior}{0.8}

\newcommand{\muv}{$\mu_v$}
\newcommand{\sigmav}{$\sigma_v$}
\newcommand{\frot}{$f_{\rm rot}$}
\newcommand{\sigmadisk}{$\sigma_{\rm disk}$}
\newcommand{\vHIrot}{$V_{\rm HI,rot}$}
\newcommand{\Rdisk}{$R_{\rm disk}$}

\newcommand{\fracdwarf}{12\%}
\newcommand{\muvdwarf}{$-195.8^{+2.0}_{-4.9}$}
\newcommand{\sigmavdwarf}{$28.3^{+2.0}_{-4.5}$}
\newcommand{\dlogzthreevsfive}{$-4.4$}

\newcommand{\risokpc}{0.56} 
\newcommand{\risoas}{150''}

\newcommand{\dlogzrgbmw}{$-2.5$}
\newcommand{\dlogzrgbgss}{$-1.3$}
\newcommand{\dlogzagbfourdvstwod}{$-16.4$}
\newcommand{\dlogzagbdwarf}{$-6.9$}

\newcommand{\dlogzagbgss}{$-8.4$}
\newcommand{\sigmargbsigmadisk}{1.8}

\newcommand{\vloshalorot}{$-317.7 \pm 9.6$}
\newcommand{\vloshalorotmin}{$-348$}
\newcommand{\vloshalorotmax}{$-303$}
\newcommand{\dlogztwovsonediskoffsethalo}{10.8}
\newcommand{\dlogztwodiskrotvsoffsethalo}{11.6}
\newcommand{\sigmathick}{100}
\newcommand{\sigmathin}{30}
\newcommand{\frotthin}{1.0}
\newcommand{\frotthick}{0.9}
\newcommand{\frothalo}{0.26}
\newcommand{\sigmahalorot}{155}
\newcommand{\dlogztwodiskoffsetvsonediskrot}{6.5}

\newcommand{\mbh}{$M_{\rm BH} \sim 2-4 \times 10^6 M_\odot$}
\newcommand{\vmaxrot}{46}
\newcommand{\kvpr}{$\mathcal{N}(10,4.4)$}

\newcommand{\krinner}{$7.6^{+1.4}_{-3.7}$}
\newcommand{\thetarinnerval}{$-9.7^{+15.2}_{-39.4}$}
\newcommand{\krouter}{$3.5^{+1.3}_{-2.8}$}
\newcommand{\thetarouterval}{$-135.6^{+35.5}_{-95.6}$}

\newcommand{\thetarinner}{$-10$}
\newcommand{\thetarouter}{$-135$}
\newcommand{\thetardisk}{$-55$}

\newcommand{\kvouter}{$3.4^{+0.9}_{-2.5}$}
\newcommand{\thetavouter}{$-131.5^{+33.6}_{-61.7}$}
\newcommand{\kvinner}{$9.8^{+2.0}_{-4.5}$}
\newcommand{\thetavinner}{$-8.5^{+22.9}_{-57.8}$}

\newcommand{\fracdiskhaloinsample}{83\%}
\newcommand{\thetardiskval}{$-54.9^{+0.2}_{-0.6}$}
\newcommand{\krdisk}{$88.9^{+0.4}_{-0.9}$}

\newcommand{\meaninner}{$-193.1 \pm 0.04$}
\newcommand{\stdinner}{$27.0^{+2.4}_{-0.4}$}
\newcommand{\gammainner}{$0.12^{+0.03}_{-0.05}$}
\newcommand{\kappainner}{$-0.26^{+0.11}_{-0.08}$}

\newcommand{\meanouter}{$-194.1 \pm 0.05$}
\newcommand{\stdouter}{$32.3^{+1.1}_{-0.4}$}
\newcommand{\gammaouter}{$0.11^{+0.03}_{-0.04}$}
\newcommand{\kappaouter}{$-0.06^{+0.07}_{-0.06}$}

\newcommand{\fracunbound}{4.5}
\newcommand{\fracunboundsb}{5.7}
\newcommand{\fracunboundpos}{3.6}
\newcommand{\fracunboundhow}{5.0}

\newcommand{\fehphotdwarfvel}{$-0.60 \pm 0.006$}
\newcommand{\sigmafehphotdwarfvel}{$0.44^{+0.005}_{-0.006}$}
\newcommand{\fehphotdwarfsb}{$-0.59 \pm 0.006$}
\newcommand{\sigmafehphotdwarfsb}{$0.41 \pm 0.006$}
\newcommand{\fehphotgss}{$-0.5$}

\newcommand{\pabreak}{$-25$}
\newcommand{\choiorbitpa}{65}
\usepackage{threeparttable,hyperref}
\usepackage{amsmath,physics,mathrsfs,amssymb}

\received{January 17, 2025}
\revised{July 14, 2025}
\accepted{July 29, 2025}

\shorttitle{M32 Kinematical Modeling}
\shortauthors{Escala et al.}

\graphicspath{{./}{figures/}}

\begin{document}

\title{Kinematical Modeling of the Resolved Stellar Outskirts of M32: Constraints on Tidal Stripping Scenarios}

\correspondingauthor{Ivanna Escala}
\email{iescala@stsci.edu}

\author[0000-0002-9933-9551]{Ivanna Escala}
\affiliation{Space Telescope Science Institute, 3700 San Martin Drive, Baltimore, MD 21218, USA}
\affiliation{Department of Astrophysical Sciences, Princeton University, 4 Ivy Lane, Princeton, NJ 08544, USA}
\affiliation{The Observatories of the Carnegie Institution for Science, 813 Santa Barbara St, Pasadena, CA 91101, USA}

\author[0000-0001-8275-9181]{Douglas Grion Filho}
\affiliation{Department of Astronomy \& Astrophysics, University of California Santa Cruz, 1156 High Street, Santa Cruz, CA 95064, USA}

\author[0000-0001-8867-4234]{Puragra Guhathakurta}
\affiliation{UCO/Lick Observatory, Department of Astronomy \& Astrophysics, University of California Santa Cruz, 1156 High Street, Santa Cruz, CA 95064, USA}

\author[0000-0003-0394-8377]{Karoline M. Gilbert}
\affiliation{Space Telescope Science Institute, 3700 San Martin Drive, Baltimore, MD 21218, USA}
\affiliation{The William H. Miller III Department of Physics \& Astronomy, Bloomberg Center for Physics and Astronomy, Johns Hopkins University, 3400 N. Charles Street, Baltimore, MD 21218, USA}

\author[0000-0003-4207-3788]{Mark Fardal}
\affiliation{Eureka Scientific}

\author[0000-0001-8536-0547]{L.~R. Cullinane}
\affiliation{Leibniz-Institut f{\"u}r Astrophysik (AIP), An der Sternwarte 16, D-14482 Potsdam, Germany}
\affiliation{The William H. Miller III Department of Physics \& Astronomy, Bloomberg Center for Physics and Astronomy, Johns Hopkins University, 3400 N. Charles Street, Baltimore, MD 21218, USA}

\author[0000-0002-9599-310X]{Erik Tollerud}
\affiliation{Space Telescope Science Institute, 3700 San Martin Drive, Baltimore, MD 21218, USA}

\author[0000-0001-8481-2660]{Amanda C.~N. Quirk}
\affiliation{Department of Astronomy, Columbia University, 538 West 120th Street, New York, NY 10027, USA}

\author[0000-0002-3038-3896]{Zhuo Chen}
\affiliation{Department of Astronomy, University of Washington, Box 351580, Seattle, WA 98195, USA}

\author{Molly Hyver}
\affiliation{Los Altos High School, 201 Almond Ave, Los Altos, CA 94022, USA}


\author[0000-0002-7502-0597]{Benjamin F. Williams}
\affiliation{Department of Astronomy, University of Washington, Box 351580, Seattle, WA 98195, USA}



\begin{abstract}
As the only compact elliptical close enough to resolve into individual stars, the satellite dwarf galaxy M32 provides a unique opportunity for exploring the origins of such rare galaxies. 
In this work, we combined archival and novel Keck/DEIMOS spectroscopy from a southern extension of the Spectroscopic and Photometric Landscape of Andromeda's Stellar Halo (SPLASH) survey with optical HST imaging from the Panchromatic Hubble Andromeda Southern Treasury (PHAST) survey. The resulting sample of \ngiants\ giant stars is unprecedented both in size and spatial coverage (\rammin--\rammax\ arcmin, or out to $\sim$\rmajeff\reff\ and $\sim$\rmineff\reff\ along M32's major and minor axes) for probing the resolved stellar outskirts of M32. Given the structurally complex region near M32 on the sky, we modeled M32's line-of-sight kinematics simultaneously alongside M31's rotating stellar disk and potential outliers corresponding to M31's kinematically hot stellar halo and/or tidal substructure. 
Inside the radius corresponding to the observed twisting of isophotal contours in M32's surface brightness profile ($R_{\rm iso} \sim$ 5\reff $\sim$ 150'' or 0.56 kpc), M32 exhibits a line-of-sight velocity distribution characteristic of 
ordered rotation, transitioning to 
a distribution with heavier outliers beyond this radius. Within $R_{\rm iso}$, the rotational direction is aligned with M32's major-axis rotation, but 
shifts to become roughly aligned with M32's minor axis beyond $R_{\rm iso}$. 
We interpret these kinematical signatures in the stellar outskirts of M32 as evidence of tidal distortion from interactions with M31 and discuss their implications for M32 formation pathways.

\end{abstract}


\section{Introduction} \label{sec:intro}

The origins of rare compact dwarf elliptical galaxies (cEs), which lie at the relatively low-mass end of the early-type galaxy population (8 $< \log(M_\star/M_\odot) <$ 10), have long presented a challenge to our understanding of galaxy formation \citep{Mieske2005,SmithCastelli2008,Chilingarian2007,Chilingarian2009,Price2009,Norris2014,Chilingarian2015,FerreMateu2018,FerreMateu2021,Caso2024}. With only approximately 200 known cEs in the local universe, these galaxies typically have effective radii between 0.1--0.7 kpc, which is smaller than their low-density dwarf elliptical counterparts at a given stellar mass. Although cEs appear to follow some typical galaxy scaling relations \citep{Kormendy2009,Paudel2014}, they tend to be significant outliers on the stellar mass--metallicity relation for early-type galaxies (e.g., \citealt{Gallazzi2005,Kirby2013}).

Moreover, cEs have been found in a variety of environments (e.g., \citealt{Norris2014}), ranging from those clearly interacting with a massive host galaxy \citep{Huxor2011} to those that appear to be isolated \citep{Huxor2013,Paudel2014}. Metal-rich cEs near a host may be the remnants of tidally stripped massive progenitors (\citealt{Bekki2001,Khoperskov2023}, but see also \citealt{Du2019,Bian2024}), whereas isolated cEs could either have been ejected from their original galactic systems by gravitational interactions \citep{Chilingarian2015} or have formed in an intrinsically compact state \citep{Burkert1994,Kormendy2009}. Recent observational studies support a mix of formation pathways for cEs (e.g., \citealt{FerreMateu2018,FerreMateu2021}), although their relative contributions still warrant exploration in a cosmological context \citep{Deeley2023,Jang2024,Bian2024}.

\begin{figure*}
    \centering
    \includegraphics[width=\textwidth]{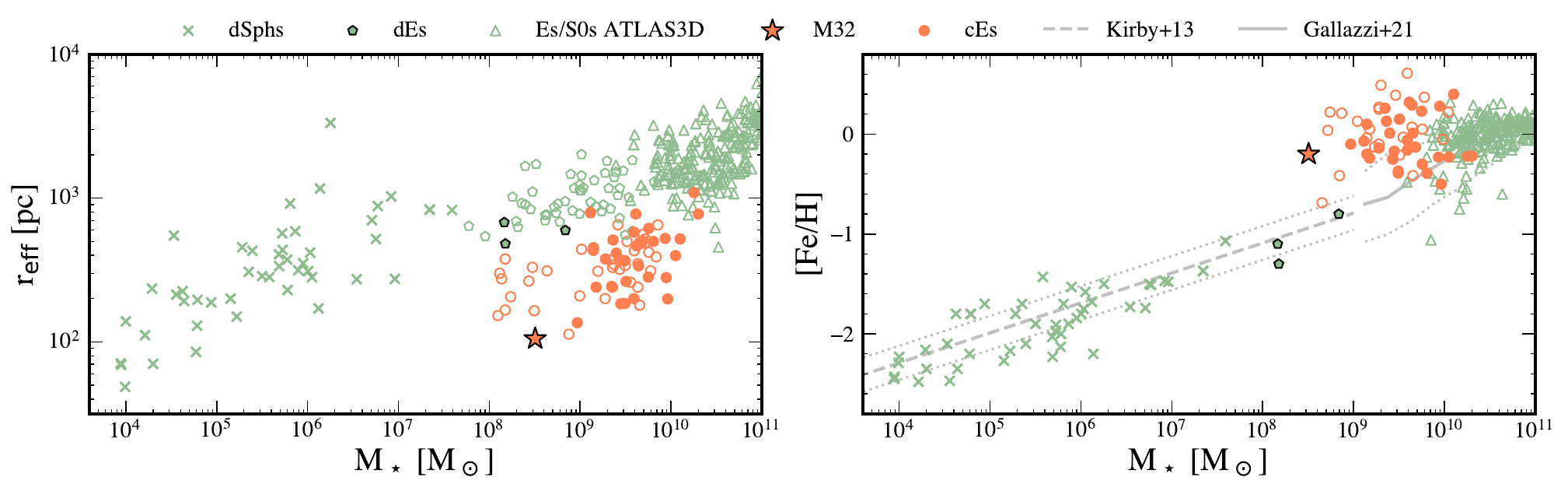}
    \caption{The position of M32 (orange outlined star) in the galaxy stellar mass--size relation (left) and stellar mass--metallicity relation (right). See main text for the adopted values for M32 (Section~\ref{sec:intro}). For context, we show data for different galaxy types: compact ellipticals (cEs) from \citet{FerreMateu2018,FerreMateu2021} (filled orange circles) and the AIMSS compilation for compact stellar systems with $0.1<r_{\rm eff} \ ({\rm kpc})<0.9$ and $8<\log_{10}(M_\star/M_\odot)<11$ (\citealt{Norris2014}, \citealt{Janz2016}, and specifically including \citealt{Guerou2015}; open orange circles); Local Group dwarf spheroidals selected from the Local Volume Database \citep{Pace2024}, largely based on data compiled by \citet{McConnachie2012} (green crosses); similarly selected Local Group dwarf ellipticals (green outlined pentagons) and AIMSS dwarf ellipticals (\citealt{Norris2014}; open green pentagons); and massive early type elliptical (E) and lenticular (S0) galaxies from ATLAS$^{\rm 3D}$ (\citealt{Cappellari2013,McDermid2015}; green open triangles). We also show the stellar mass--metallicity relation for Local Group dwarf galaxies (\citealt{Kirby2013}; dashed grey line) and massive galaxies from SDSS (\citealt{Gallazzi2021}; solid grey line), including rms dispersions (dotted grey lines).
    }
    \label{fig:relations}
\end{figure*}

The nearby compact dwarf elliptical satellite galaxy, M32, is the prototype for the cE class.
As the only cE close enough to resolve into individual stars, M32 provides a unique opportunity for exploring cE origins. Even among cEs, M32 has extreme properties (\added{see Figure~\ref{fig:relations})}: it is highly compact (\reff\ $\sim$ 0.1 kpc; \citealt{Choi2002}) and a 
significant outlier on the stellar mass--metallicity relation ([Fe/H] $\sim -0.2$; \citealt{Monachesi2011}). The stellar mass and dynamical mass within \reff\ of M32 is approximately $M_{\star} \sim 3.2 \times 10^8 M_\odot$ and $5.2 \times 10^8 M_\odot$ respectively \citep{Bender1996,McConnachie2012}, where M32 is known to possess a central supermassive black hole (\mbh; \citealt{Joseph2001,vanderMarel1997a,Verolme2002,vandenBosch2010}). Integrated light studies suggest the presence of a central population of young metal-rich stars in M32 \citep{Schiavon2004,Rose2005,Coelho2009}, where resolved stellar photometry also shows evidence for a substantial intermediate-age population \citep{Monachesi2011,Monachesi2012,Jones2023}.

Owing to its proximity to M31 ($D_{\rm M31} = 20.6^{+13}_{-21}$ kpc; \citealt{Savino2022}), tidal interactions may provide a plausible formation pathway for M32 that could explain its unusual properties.
M32 could be the remnant bulge of a spiral galaxy that was tidally stripped inside M31's potential \citep{Bekki2001}, or alternatively, M32 may have originally been a dwarf galaxy that was compacted over a series of interactions with M31, but without experiencing significant mass loss \citep{Du2019}.  In particular, the former hypothesis has long been an explanation for M32's origin based on its compact nature and its metal-rich intermediate-age stellar populations. 

Upon discovery of M31's Giant Stellar Stream (GSS), \citealt{Ibata2001} proposed M32 as a possible candidate for the stream's progenitor owing its clear status as a bright and close companion of M31. \citealt{Williams2015} further speculated that M32 may be the remnant of a massive, gas-rich galaxy that strongly interacted with M31 2--4 Gyr ago, triggering similarly timed starbursts both in M31's disk and M32 \citep{Monachesi2012,Bernard2015a}.  More recently, \citealt{DSouzaBell2018} 
argued that M32 may be the core of an M33- or LMC-mass spiral galaxy based on the observed properties of M31's massive, metal-rich, and intermediate-age stellar halo, which point to a single dominant merger when compared to cosmological simulations. This scenario could simultaneously explain M31's system of major tidal features, which includes the GSS, and the observed properties of both M31's disk and M32.

Numerical simulations of the evolution of M31-like systems have provided potential clues about the origins of M32.
Early N-body simulations of interactions between spherical or elliptical progenitors and an M31-like host \citep{Choi2002,Johnston2002} concluded that M32 was unlikely to have experienced significant tidal evolution (e.g., \citealt{Penarrubia2008}), but could not rule out the bulge of a spiral galaxy as its precursor. 
Subsequent modeling tailored to the M31-M32 interaction focused on scenarios involving recent ($\lesssim$800 Myr) off-center collisions between M32 and M31's disk in an attempt to reproduce M31's star-forming rings, including the split in the 10 kpc ring near M32 on the sky \citep{Gordon2006,Block2006,Dierickx2014}. However, the long-term stability of the 10 kpc star-forming ring on $\gtrsim$500 Myr timescales \citep{Lewis2015} largely rules out a recent collision between M32 and M31's disk in the context of existing models, at least as an explanation for M31's young stellar and gaseous features (c.f.\ \citealt{Davidge2012}).  For example, simulations of a major merger in an M31-like system with an $M_\star \sim 10^{10} M_\odot$ progenitor occurring 2--3 Gyr ago can reproduce M31's star-forming ring without invoking an M32-like galaxy \citep{Hammer2010,Hammer2018}.
Moreover, the first and only self-consistent model of the M31-M32 interaction found tidal stripping of an initially massive M32 ($M_{\rm tot} \sim 10^{10} M_{\odot}$) over a single pericentric passage to be insufficient in terms of producing a compact elliptical morphology, thereby concluding that cEs like M32 may be intrinsically compact  \citep{Dierickx2014}.

Existing observational measurements suggest minimal direct evidence for tidal stripping within M32's central $\sim$1 kpc, at face value supporting an intrinsically compact origin. Despite the presence of elongated isophotes in M32's outer surface brightness profile, 
\citealt{Choi2002}, hereafter C02, estimated that M32's effective radius experienced little evolution with time and therefore was not influenced by tidal distortions.
However, \citealt{Graham2002} argued that the surface brightness profile over the full radial range probed was better explained by a S\'{e}rsic bulge and ``fossil'' exponential disk, which corresponds to M32's observed excess light at large radii, than the elliptical de Vaucouleurs profile of C02.  The only resolved stellar kinematical study of M32's outskirts to date (\citealt{Howley2013}, hereafter H13) found little evidence of tidal distortion within $\sim$8\reff, with no detected change in kinematics across the region where M32's isophotal elongation begins (C02). In addition, there is no clear observational connection between M32 and tidal debris in M31's stellar halo that may correspond to stripped material from M32's putative progenitor, where M32's present-day position is inconsistent with current predictions from N-body models for the location of the GSS progenitor core on the eastern side of M31 \citep{Fardal2013,Escala2022}.

In this work, we revisit M32's resolved stellar kinematics with the goal of placing updated  observational constraints on M32's velocity structure at large radii 
to inform its origins as a cE galaxy, particularly scenarios involving strong tidal interactions with M31.  Resolved kinematical studies of M32 are lacking owing to M32's steep surface brightness profile combined with its projected location at the edge of M31's disk, such that observations must contend with both stellar crowding and contamination from M31.
To circumvent this, we utilize an expanded dataset in sample size and areal coverage, combining all available ground-based Keck/DEIMOS stellar spectroscopy in the vicinity of M32 from southern extensions of the Spectroscopic and Photometric Landscape of Andromeda's Stellar Halo survey (SPLASH; e.g., \citealt{Guhathakurta2005,Gilbert2006,Gilbert2012,Tollerud2012}) with high-precision Hubble Space Telescope (HST) photometry from the Panchromatic Hubble Andromeda Southern Treasury  (PHAST; Chen et al.\ 2025, accepted). PHAST is a new legacy survey imaging the southern half of M31's star-forming disk at optical and near-ultraviolet wavelengths, covering key diagnostic features sensitive to M31's complex merger history, specifically including intersections with M32.

We summarize the observations and construction of the full spectroscopic dataset in Section~\ref{sec:obs} and identify giant stars in M32 and M31 from the sample in Section~\ref{sec:mem}.  In Section~\ref{sec:kinematics}, we apply a spatially continuous Bayesian mixture model to describe the line-of-sight velocity distribution of the structurally complex region near M32 to isolate the kinematics of the dwarf galaxy relative to stellar populations in M31. We present M32's velocity structure with respect to 
its ``inner'' and ``outer'' regions as defined by its surface brightness profile
in Section~\ref{sec:m32} and discuss implications for M32's origin in Section~\ref{sec:discuss}. Throughout this work, we assume that M32 and M31 are located at the same distance ($D_{\rm M31} = 776.2^{+22}_{-21}$ kpc; \citealt{Savino2022}).

\section{Spectroscopic Data}
\label{sec:obs}

\begin{figure*}
    \centering
    \includegraphics[width=\textwidth]{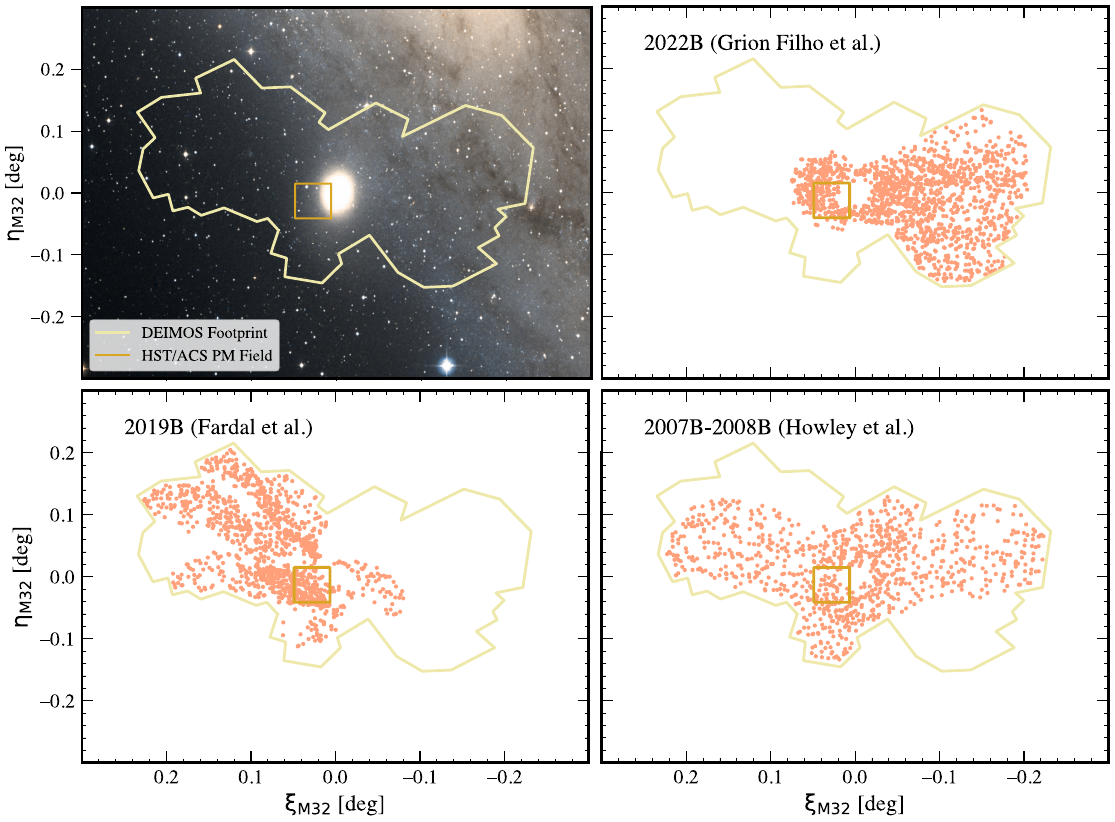}
    \caption{(Upper left) Approximate footprint of Keck/DEIMOS spectroscopic fields (yellow outline) 
    in the vicinity of M32 used in this work (Section~\ref{sec:obs}; Table~\ref{tab:obs}), overlaid on a DSS image of M32 and M31. We also show the location of an archival HST/ACS field targeting M32 (orange rectangle; HST GO 9392 and 15658) suitable for proper motion measurements (Fardal et al., in preparation; Patel et al., in preparation). (Upper right and bottom panels) Sky location of spectroscopic targets with successfully extracted 1D spectra (excluding serendipitous detections), separated according to observing program (2022B, 2019B, H13). The central regions are inaccessible for ground-based resolved stellar spectroscopy due to crowding.
    }
    \label{fig:obs}
\end{figure*}

We combined previously unpublished and archival resolved stellar spectroscopy obtained with the DEIMOS instrument on Keck II to produce a sample of \ntargets\ stars with successful velocity measurements unprecedented both in size and spatial coverage of the stellar outskirts of M32. Table~\ref{tab:obs} summarizes each observational program and Figure~\ref{fig:obs} shows the approximate footprint of the combined set of 20 DEIMOS slitmasks, in addition to targets with successfully extracted 1D spectra for each program. In Sections~\ref{sec:2022B} and~\ref{sec:2019B}, we describe the program-specific observational details for unpublished spectroscopy from campaigns in Fall 2022 (Grion Filho et al., in preparation) and Fall 2019 (Fardal et al., in preparation) respectively. Section~\ref{sec:howley} outlines the spectroscopy from Fall 2007 and 2008 published by H13. When constructing the combined spectroscopic sample of unique targets, we flagged duplicates present across each individual sample, giving preference to the 2022B, 2019B, and H13 samples in that order.

For all observations, the 1200 $\ell$/mm DEIMOS grating was used with the OG550 order blocking filter and a central wavelength of 7800 \AA\ for the science configuration, resulting in a spectral dispersion of 0.33 \AA\ per pixel, with $\sim$1 hr science exposures. The two-dimensional and one-dimensional spectra were reduced using versions of the {\tt\string spec2d} and {\tt\string spec1d} pipelines \citep{Cooper2012,Newman2013} modified for stellar sources \citep{SimonGeha2007}. Radial velocities were obtained by cross-correlating the one-dimensional spectra against empirical templates of hot stars, including A-band corrections for slit miscentering \citep{Sohn2007} and heliocentric corrections. \added{We applied a statistical, as opposed to slit-by-slit, A-band correction as a function of the target's position on each slitmask computed from a polynomial function fit to the individual A-band velocities of targets with the highest quality measurements ($Q=4$, see below; \citealt{Gilbert2022,Quirk2022}). We removed targets with unreliable velocity corrections, in which the statistical A-band correction deviates from the median value for the associated slitmask by $>$3$\sigma$, from the spectroscopic sample.
}

\added{Following H13}, the total velocity uncertainty ($\delta v_{\rm tot}$) was determined 
\added{from both random and systematic contributions,
\begin{equation}
    \delta v_{\rm tot} = \sqrt{ (1.85 \times \delta v_{\rm cc})^2 + \delta v_{\rm sys}^2},
\end{equation}
where $\delta_{\rm cc}$ is the cross-correlation based random uncertainty from \texttt{zspec} and $\delta v_{\rm sys}$ is the systematic velocity uncertainty of 2.2 \kms\ from \citet{SimonGeha2007}. The random uncertainty is multiplied by a scale factor determined from a sample of duplicate M31 RGB stars to account for the fact that $\delta v_{\rm cc}$ is likely an underestimate compared to the random uncertainties calculated using a Monte Carlo approach by \citeauthor{SimonGeha2007}. 
}

To assess the reliability of the \added{cross-correlation based} velocity measurements, the raw two-
dimensional spectra, extracted one-dimensional spectra,
and best-fit empirical templates were visually inspected
in the {\tt\string zspec} software (D. Madgwick, DEEP2 survey) and
assigned a quality code (Q; e.g., \citealt{Guhathakurta2006}). We restrict our analysis to objects with successful radial velocity measurements, which match at least one strong spectral feature ($Q = 3$ or $Q = 4$). The median total velocity uncertainty across the entire sample is \vtotmed\ \kms\ given a median sample signal-to-noise ratio (S/N) of \snmed\ per pixel.

\begin{table*}
    \centering
    \begin{threeparttable}
    \caption{Summary of Keck/DEIMOS Observations in the Vicinity of M32}
    \begin{tabular*}{\textwidth}{@{\extracolsep{\fill}}lccrccccc}
        \hline \hline
        Slitmask & R.A. & Decl. &  \multicolumn{1}{p{1.5cm}}{\centering Mask P.A.\\ ($^\circ$ E of N)} & \multicolumn{1}{p{1.5cm}}{\centering Obs. Date\\(UT)} &  \multicolumn{1}{p{1.3cm}}{\centering t$_{\rm exp}$\\(min)} & Airmass & \multicolumn{1}{p{1.2cm}}{\centering No.\\Targets\tnote{a}} & \multicolumn{1}{p{1.0cm}}{\centering No.\\RV\tnote{b}} \\ \hline
        \multicolumn{9}{c}{2022B (Grion Filho et al.)} \\ \hline
        M32RA1 & 00:42:25.56 & +40:54:20.3 & $-$186.2 & 2022 Sep 27 & 60.0 & 1.07 & 175 & 173\\
        M32RB1 & 00:42:19.34 & +40:54:13.8 & $-$237.4 & 2022 Sep 28 & 60.0 & 1.18 & 175 & 169\\
        M32RB2 & 00:42:19.57 & +40:54:08.6 & $-$226.0 & 2022 Sep 27 & 60.0 & 1.13 & 166 & 157\\
        M32RC1 & 00:42:20.52 & +40:52:39.8 & $-$99.4 & 2022 Sep 27 & 60.0 & 1.52 & 184 & 177\\
        M32RC2 & 00:42:21.09 & +40:52:38.9 & +111.6 & 2022 Sep 27 & 76.8 & 1.28 & 176 & 156\\
        M32RD1 & 00:42:23.35 & +40:51:18.2 & $-$116.0 & 2022 Sep 27 & 60.0 & 1.23 & 177 & 171\\
        M32RD2 & 00:42:23.76 & +40:51:18.4 & $-$111.4 & 2022 Sep 28 & 60.0 & 1.28 & 172 & 160\\
        M32RE1 & 00:42:21.55 & +40:49:40.8 & $-$133.4 & 2022 Sep 28 & 54.0 & 1.13 & 185 & 183\\
        M32RE2 & 00:42:21.15 & +40:49:38.6 & $-$143.8 & 2022 Sep 27 & 60.0 & 1.10 & 179 & 174\\ \hline
        \multicolumn{9}{c}{2019B (Fardal et al.)} \\ \hline
        m32p2a & 00:43:03.14 & +40:56:28.6 & $-$177.0 & 2019 Oct 23 & 67.3 & 1.07 & 214 & 103\\
        m32p2b & 00:43:02.86 & +40:56:30.6 & $-$115.8 & 2019 Oct 23 & 95.0 & 1.23 & 212 & 111\\
        m32p3a & 00:43:18.52 & +40:54:38.2 & $-$110.4 & 2019 Oct 22 & 95.0 & 1.29 & 216 & 90\\
        m32p3b & 00:43:18.66 & +40:54:35.3 & $-$89.6 & 2019 Oct 22 & 88.3 & 1.96 & 213 & 107\\
        m32pm1 & 00:43:00.21 & +40:53:28.4 & $-$158.2 & 2019 Oct 22 & 85.1 & 1.08 & 223 & 126\\
        m32pm4 & 00:43:00.29 & +40:50:23.8 & $-$94.0 & 2019 Oct 23 & 80.0 & 1.71 & 210 & 111\\ \hline
        \multicolumn{9}{c}{2007B--2008B (Howley et al.)} \\ \hline
        M32\_1 & 00:42:38.55 & +40:51:29.7 & $-$112.6 & 2007 Nov 14 & 60.0 & 1.26 & 201 & 187\\
        M32\_2 & 00:43:04.03 & +40:55:07.1 & $-$95.6 & 2008 Aug 03 & 60.0 & 1.64 & 192 & 167\\
        M32\_3 & 00:43:11.34 & +40:52:41.7 & $-$110.7 & 2008 Aug 03 & 60.0 & 1.29 & 199 & 134\\
        M32\_4 & 00:42:14.59 & +40:54:39.0 & $-$121.8 & 2008 Aug 04 & 60.0 & 1.18 & 152 & 134\\
        M32\_5 & 00:42:14.50 & +40:52:11.4 & $-$152.7 & 2008 Aug 04 & 60.0 & 1.09 & 183 & 157\\
        \hline
    \end{tabular*}
    \begin{tablenotes}
    \footnotesize
    \item[a] Number of targets on the slitmask with successfully extracted one-dimensional spectra, excluding serendipitous detections.
    \item[b] Number of targets with successful radial velocity measurements (Section~\ref{sec:obs}).
    \end{tablenotes}
    \label{tab:obs}
    \end{threeparttable}
\end{table*}

\subsection{2022B (Grion Filho et al.) Observations}
\label{sec:2022B}

We observed nine slitmasks targeting M32 in a ``Japanese fan'' pattern with Keck/DEIMOS in Fall 2022 as part of a spectroscopic survey of M31's southwestern disk region (Grion Filho et al., in preparation).\footnote{We observed a tenth mask, M32RA2, which is excluded from this analysis owing to significant issues with the wavelength solution. It will be included in a forthcoming paper (Grion Filho et al., in preparation).} This survey is designed as a complement to the Spectroscopic and Photometric Landscape of Andromeda's Stellar Halo survey of M31's northeastern disk region (SPLASH; \citealt{Dorman2012,Dorman2013,Dorman2015}), which selected targets in part from the HST-based Panchromatic Hubble Andromeda Treasury (PHAT) survey \citep{Dalcanton2012,Williams2014}.

Figure~\ref{fig:obs} shows the location of spectroscopic targets on the 2022B masks in M32-centric coordinates. The fan pattern, which is anchored to the east of M32's compact center, was designed to increase the density of spectroscopic targets near M32's center while simultaneously including repeat targets on overlapping masks. The primary purpose of the repeat targets is to enhance the S/N for spectra of individual stars with higher probabilities of being associated with M32 ($>$40\%) in order to measure their detailed chemical abundances in future work (Escala et al., in preparation).\footnote{When constructing our final sample of unique targets, we selected the observation with the highest S/N spectrum per repeat target.}

\begin{figure}
    \centering
    \includegraphics[width=\columnwidth]{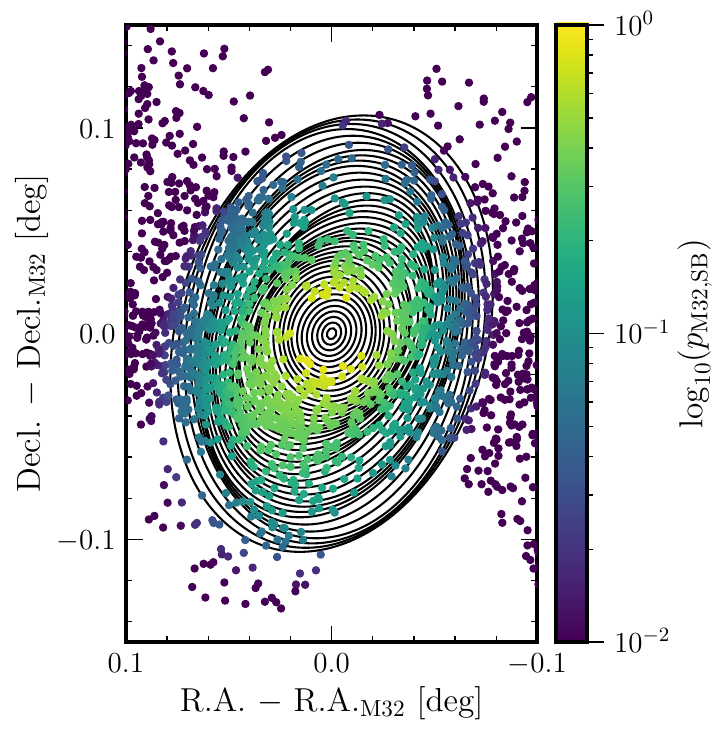}
    \caption{Location of spectroscopic targets near M32 (Table~\ref{tab:obs}; Section~\ref{sec:obs}), colored by the probability of belonging to M32 based on the relative I-band surface brightness between M32 (C02) and M31 (assuming a bulge, disk, and halo model; \citealt{Dorman2013}). We also show M32 isophotes from C02 for reference (black curves). Owing to M32's steep surface brightness profile and contamination from M31's disk to the northwest (Figure~\ref{fig:obs}), the likelihood that a given target belongs to M32 decreases rapidly with projected radius from M32's center.
    }
    \label{fig:pm32sb}
\end{figure}

We identified probable M32 stars by computing a likelihood based on the relative I-band surface brightness between M32 and M31 (\psb). Following H13, we modeled the 2D surface brightness of M32 as a series of isophotes with varying I-band surface brightness, ellipticity and position angle as a function of major/minor axis distance from M32 (C02; Figure~\ref{fig:pm32sb}). We modeled the 2D surface brightness profile of M31 as the sum of a S\'{e}rsic bulge, exponential disk, and cored power-law halo as a function of M31-centric projected radius and position angle, assuming the structural parameters derived by \citealt{Dorman2013} from combining PHAT luminosity functions with SPLASH radial velocities (see their Table~2 and Section~4.1.1). The M32 probability for a given star is thus the ratio of the predicted M32 versus M31 I-band magnitudes at its location, \psb\ $= I_{\rm M32}/(I_{\rm M31} + I_{\rm M32}$), where $I_{\rm M32} = 10^{-2\Sigma_{\rm M32}/5}$ and $\Sigma_{\rm M32}$($\alpha$, $\delta$) is the 2D surface brightness in magnitudes per square arcsecond at sky coordinates ($\alpha$, $\delta$). On average, we expect M31's bulge, disk, and halo components to contribute \percentbulge, \percentdisk, and \percenthalo\ respectively to M31's total light over the DEIMOS footprint near M32 (Figure~\ref{fig:obs}).\footnote{As discussed in Section~\ref{sec:m32}, M31's global surface brightness profile is known to substantially overpredict M31's disk fraction relative to M31's spheroid in the vicinity of M32 (see Appendix by \citealt{Dorman2012} and discussion by H13). Throughout this work, we utilize $p_{\rm M32,SB}$ only as guide on where to expect M32 stars based on the surface brightness profile alone.}

When designing the 2022B observations, we selected isolated targets from stellar catalogs based on HST/ACS photometry from PHAST (Chen et al.\ 2025, accepted). We identified crowded sources in the PHAST catalog using the empirical isolation criterion of H13, which excludes a given star as a possible spectroscopic target if the point spread function (PSF) of any of its neighbors significantly overlaps with that of the target,
\begin{equation}
    m_{\rm tgt} > m_{\rm nbr} + \frac{\theta_{\rm tgt,nbr}}{\theta_{\rm DEIMOS}} - m_{\rm lim},
\label{eq:crowd}
\end{equation}
where $m_{\rm tgt}$ and $m_{\rm nbr}$ are the F814W magnitudes of a given target and neighbor, $\theta_{\rm tgt,nbr}$ is the angular separation on the sky between the target and neighbor, $\theta_{\rm DEIMOS} = 0.8''$ is the typical seeing at Keck II, and $m_{\rm lim} = 3$ corresponds to the faintest magnitude difference for which a neighbor star may significantly contaminate the PSF of the target star. Based on this criterion, M32's central $\sim$1' ($\mu_I$ $\sim$ 18.9 mag arcsec$^{-2}$) is sufficiently crowded that PHAST giant star candidates 
are inaccessible for ground-based spectroscopy (Figure~\ref{fig:obs}).

Slitmasks designated ``1'' primarily target red giant branch (RGB) star candidates, whereas slitmasks designated ``2'' primarily target asymptotic giant branch (AGB) and red-helium burning (RHeB) star candidates at approximately the same mask placement (Table~\ref{tab:obs}).\footnote{RGB candidates that had higher surface-brightness based a priori probabilities of belonging to M32 were given higher priority as targets (for chemical abundance measurements), whereas no distinction was made among RHeB or AGB candidates based on their likelihood of being associated with M32.}
We identified RGB, AGB, and RHeB candidates based on (F475W, F814W) selection boxes in the PHAST color-magnitude diagram (CMD; Figure~\ref{fig:cmd}).\footnote{These selection boxes, which were used only for spectroscopic target selection in the 2022B slitmask design, differ from the RGB and AGB classification criteria homogeneously adopted across all observational samples (Figure~\ref{fig:cmd}; Section~\ref{sec:rgb}) Moreover, stars classified as RHeB in the 2022B mask design tend to be conservatively assigned as belonging to the MW foreground based on photometry alone in our membership determination (Section~\ref{sec:m31bayes}).} We targeted stars of different spectral types and therefore different mean stellar ages in order to perform age-dependent kinematical analyses in M31's southern disk region in future work (c.f.\@ \citealt{Dorman2015,Quirk2019}; Grion Filho et al., in preparation).
We also placed \npne\ planetary nebulae (PNe) as spectroscopic targets on the slitmasks (Bhattacharya et al.\@, in preparation), which we exclude from this analysis.

\subsection{2019B (Fardal et al.) Observations}
\label{sec:2019B}

\begin{figure*}
    \centering
    \includegraphics[width=\textwidth]{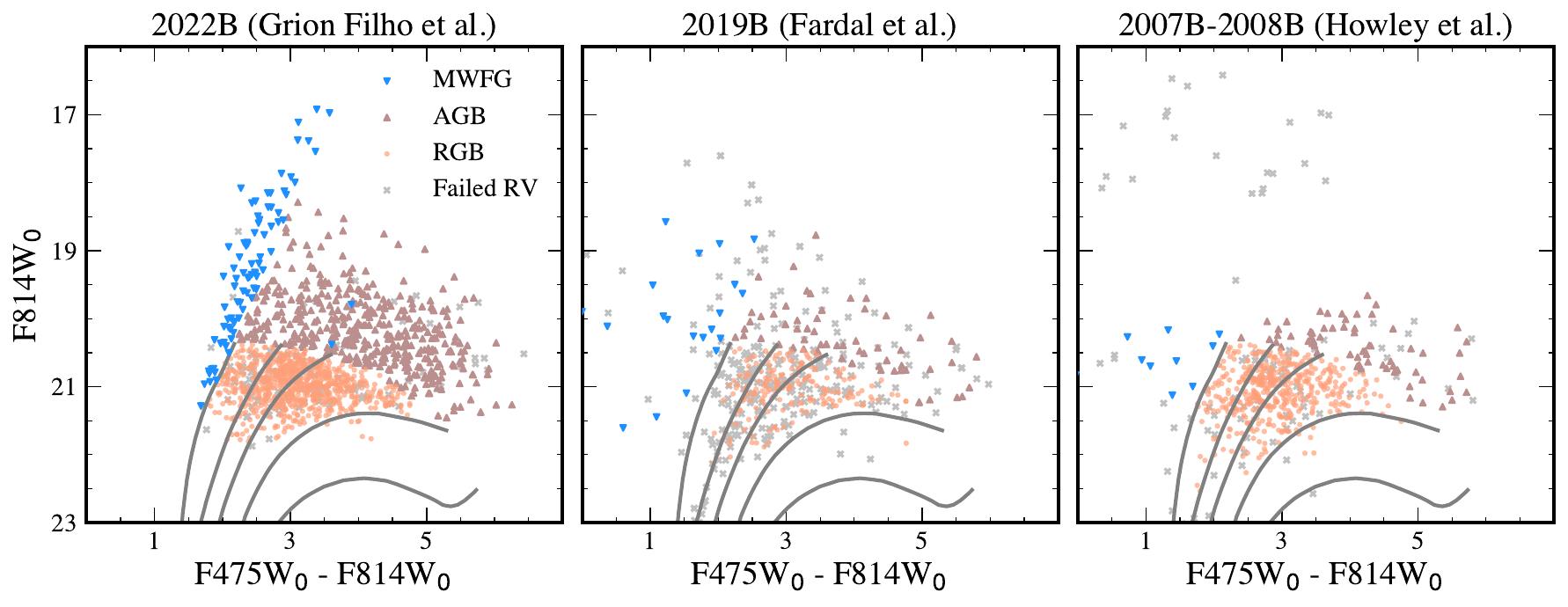}
    \caption{Extinction-corrected (F475W$_0$, F814W$_0$) PHAST (Chen et al.\ 2025, accepted) CMD for all stars with original and matched photometry separated according to observational sample (Section~\ref{sec:obs}; Table~\ref{tab:obs}). We show stars without successful velocity measurements (gray crosses), stars assigned to the MW foreground (which includes RHeB stars in M31 excluded from our sample; blue inverted triangles), and RGB (orange circles) and AGB (brown triangles) stars assigned to M31 or M32 (Section~\ref{sec:mem}). The median photometric errors are $\delta m_{\rm F475W}$ = \medblueerr\ and $\delta m_{\rm F814W}$ = \medrederr. For reference, we show PARSEC isochrones (gray lines; \citealt{Bressan2012,Tang2014,Chen2015}) assuming 12 Gyr ages and [$\alpha$/Fe] = 0 with [Fe/H] = $-$2.0, $-$1.0, $-$0.5, and 0 (from left to right).
    }
    \label{fig:cmd}
\end{figure*}

In Fall 2019, Fardal et al., in preparation observed six slitmasks targeting an HST/ACS field (HST GO 9392 and 15658) suitable for proper motion measurements near M32's core to obtain complementary line-of-sight velocity measurements for individual stars from spectroscopy. Figure~\ref{fig:obs} shows a dense clustering of targets at the location of the HST field. For this observing program, the position angle of the slitmasks increases with mask index (1--4; Table~\ref{tab:obs}), where masks labeled ``a'' and ``b'' in each pair share the same spatial position and orientation while targeting different stars.

Photometric catalogs for HST fields were obtained from the
Hubble Legacy Archive (HLA). Spectroscopic targets were selected from a combination of archival HST-based ACS F814W-band (GO 9392, PI~Mateo; \citealt{Rudenko2009}; GO 10273, PI~Crotts; \citealt{Cseresnjes2005}; GO 13691, PI Freedman; \citealt{Beaton2016}), WFPC2 F814W-band photometry (GO 8059, PI~Casertano), ACS F606W-band photometry (GO 10572, PI~Lauer; \citealt{Sarajedini2009,Monachesi2011}), and ACS F775W-band photometry (GO 9480, PI Rhodes), as well as CFHT/MegaCam $i$-band photometry from the Pan-Andromeda Archaeological Survey (PAndAS; \citealt{McConnachie2009,McConnachie2018}). 

Fardal et al.\@ used linear transformations to place the astrometry for all photometric sources on the \textit{Gaia} DR2 reference frame. The final list of possible spectroscopic targets was limited to relatively uncrowded stars using a similar isolation criterion to Eq.~\ref{eq:crowd} \citep{Dorman2012} and stars with magnitudes $-3 \lesssim m_{\rm TRGB} - m \lesssim +1.5$ (defined in the F814W band or similar). Targets selected from HST photometry were prioritized over those selected from the ground-based PAndAS survey owing to the higher likelihood of blending in the CFHT images.

Given that the available photometry is limited to a single band per source program and that a mix of filters was used for selecting spectroscopic targets, we matched targets with successfully extracted 1D spectra to (F475W, F814W) photometry from PHAST. We restricted potential matches to stars in the PHAST catalog meeting the ``good star'' criteria from \citealt{Williams2021} based on DOLPHOT \citep{Dolphin2000,Dolphin2016} quality flags in each band. For each spectroscopic target, we searched for matches within a 0.5'' radius. If \added{source} F814W photometry was unavailable for a given spectroscopic target, we selected the PHAST target with the brightest F814W magnitude as the match. Alternately, if \added{source} F814W photometry was available, we required that the magnitude of the match agreed with that of the spectroscopic target within 2$\sigma$ of the PHAST photometric errors. Figure~\ref{fig:cmd} shows the CMD for all stars with matched PHAST photometry in the 2019B sample. \added{The relative proportion of usable velocity measurements in this  sample is lower than in the 2022B and H13 samples due to targets with fainter magnitudes and low S/N $\sim$ 0. 
These targets had velocities assigned by visual inspection rather than by the \texttt{zspec} cross-correlation routine, thus we excluded these measurements from the subsequent analysis owing to concerns about their reliability.}

\subsection{2007B--2008B (Howley et al.) Observations}
\label{sec:howley}

We used archival Keck/DEIMOS spectroscopy of five slitmasks targeting M32 published by H13. Figure~\ref{fig:obs} shows the H13\ targets with successfully extracted 1D spectra, where a single slitmask is oriented along M32's major axis (P.A.\ $\sim$ $-22$ deg E of N)
and the remaining slitmasks are positioned to optimize coverage of M32's outer regions where isophotal distortion is present ($R_{\rm M32} > 150'' \sim 5$\reff; C02; Figure~\ref{fig:pm32sb}), in addition to M31's disk and halo \citep{Dorman2012}. 

H13\ selected spectroscopic targets from archival single-band CFHT/MegaCam $i'$ photometry (obtained at 0.8'' seeing) in the magnitude range $i' = 20-21.5$. They rejected targets that were affected by crowding (Eq.~\ref{eq:crowd}) or blending, where the latter effect was assessed based on visual inspection of the PSF in the MegaCam images. Given that the seeing (0.6'') during spectroscopic observations exceeded that of the photometry used in the mask design, H13\ performed further de-blending in the spatial and spectral domains by identifying serendipitously detected sources. 

Similar to the case of the 2019B observations (Section~\ref{sec:2019B}), we matched the H13\ targets to stars in the PHAST stellar catalog based solely on their spatial position, adopting the (F475W, F814W) photometry of the brightest star within the $0\farcs5$ search radius. Figure~\ref{fig:cmd} shows the PHAST CMD for all targets with matched photometry, where a number of bright targets ($m_{\rm F814W} < 19$) with failed radial velocity measurements likely reflects the presence of crowded/blended sources selected as spectroscopic targets owing to the limitations of the MegaCam imaging.

\section{M31 \& M32 Membership Determination}
\label{sec:mem}

In this section, we evaluate the spectroscopic and photometric properties of the stellar sample from Section~\ref{sec:obs} to associate stars with M31 or M32 (as opposed to the MW foreground; Section~\ref{sec:m31bayes}). We also separate the sample into populations based on stellar type (Section~\ref{sec:rgb}).

\subsection{Identifying M31/M32 Giant Stars}
\label{sec:m31bayes}

\begin{figure}
    \centering
    \includegraphics[width=\columnwidth]{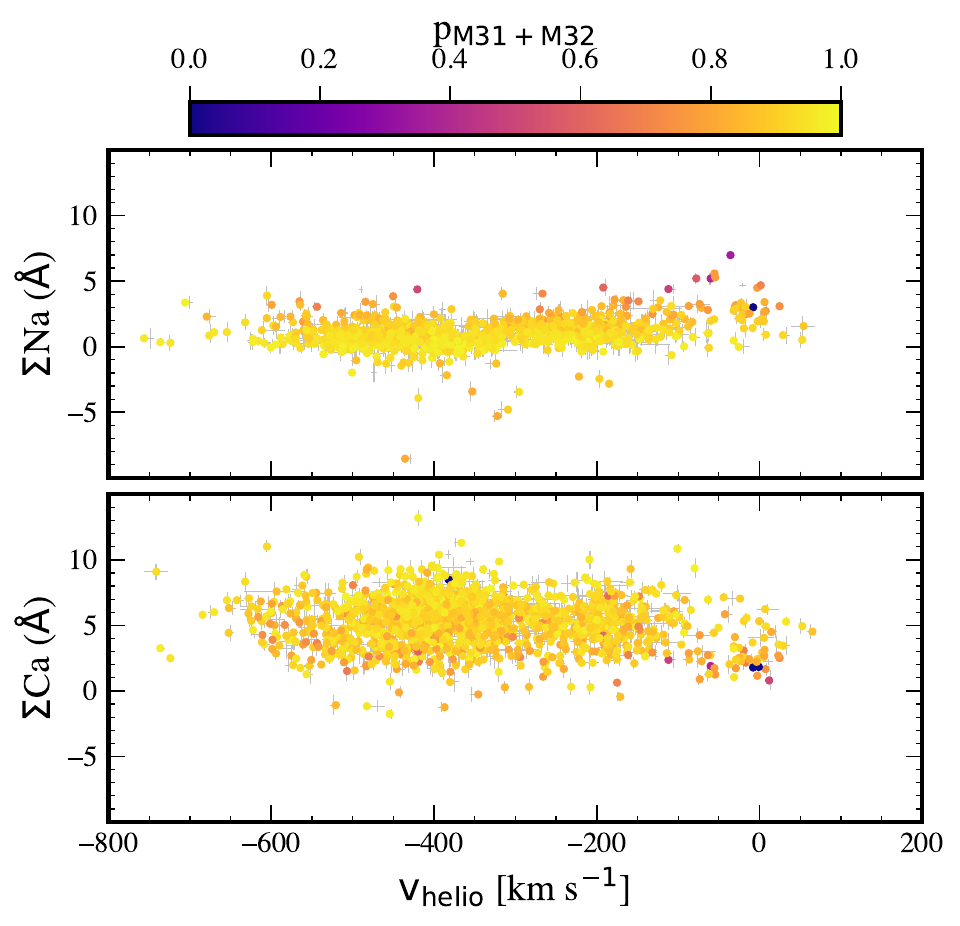}
    \caption{(Left) Summed equivalent width of the Na I $\lambda$8190 doublet ($\Sigma$Na) and (right) summed equivalent width of the Ca II triplet near $\lambda$8500  ($\Sigma$Ca) following \citet{Rutledge1997} versus heliocentric velocity ($v_{\rm helio}$). We show all starts with successful velocity measurements and equivalent width measurements with $\delta\Sigma$Na $<$ 0.8 \AA\ or $\delta\Sigma$Ca $<$ 0.8 \AA\ for clarity (Section~\ref{sec:m31bayes}). Each star is color-coded by the probability that it belongs to M31 or M32 versus the MW foreground ($p_{\rm M31+M32}$) based on population diagnostics ($\Sigma$EW, $\Sigma$Ca, and CMD position; \citealt{Escala2020b}). Most stars have a high probability of being associated with M31 or M32.
    }
    \label{fig:mem}
\end{figure}

We assigned stars to the MW foreground based on the equivalent width of the surface-gravity sensitive Na I $\lambda$8190 \AA\ doublet (\signa), the strength of the metallicity-sensitive Ca II triplet absorption feature near 8500 \AA\ (\sigca), and PHAST CMD position. We used these diagnostics in the context of the model developed by \citealt{Escala2020b}, which relies on Bayesian inference to assign each target star a probability of M31 versus MW membership based on the observed properties of over a thousand M31 and MW stars securely identified as part of the SPLASH survey of M31's southeastern stellar halo  \citep{Gilbert2006,Gilbert2012}. Although intended for M31 halo stars, the model is useful for identifying RGB and AGB stars in M32 (and in M31's disk, which should be the dominant M31 stellar population in our dataset; Section~\ref{sec:2022B}) because M32 giant stars have low surface gravity, high metallicity, and red CMD positions similar to M31 giants and relative to the MW foreground.

In contrast to \citet{Escala2020b}, we do not use radial velocity (\vhelio) as a membership diagnostic to avoid kinematically biasing our analysis, where stars with MW-like velocities (\vhelio\ $\gtrsim -150$ \kms; \citealt{Gilbert2012}) significantly overlap with the expected velocity distribution of M32 ($\mu_v \sim -200$ \kms, $\sigma_v \sim 30$ \kms; \citealt{Dorman2012,Howley2013}). In addition, we modified the Bayesian membership model to use \sigca\ as a diagnostic instead of the calcium-triplet based spectroscopic metallicity (\fehcat), which relies on a calibration between \sigca, Johnson-Cousins V/I magnitude, and metallicity. Following \citealt{Rutledge1997},
\begin{equation}
    \Sigma{\rm Ca} = 0.5 \times {\rm EW}_{\lambda 8498} + 1.0 \times {\rm EW}_{\lambda 8542} + 0.6 \times {\rm EW}_{\lambda 8662},
\end{equation}
where EW is the equivalent width of an absorption feature at a given wavelength. Due to the dependence of \fehcat\ on \sigca, the latter diagnostic similarly demonstrates a clear difference in the distributions corresponding to secure M31 (MW) stars in SPLASH: $\langle\Sigma$Ca$\rangle$ = 5.60 (3.00) \AA\ and $\langle\delta\Sigma$Ca$\rangle$ = 1.71 (0.88) \AA. The advantages of using \sigca\ instead of \fehcat\ are twofold given that \fehcat\ calibrations depend on apparent magnitude: the reliance on transformations between the ACS and Johnson-Cousins photometric systems (e.g.\ \citealt{Sirianni2005}) is eliminated and information from \sigca\ remains usable even in the absence of PHAST-matched photometry (Section~\ref{sec:2019B},~\ref{sec:howley}).

We parameterized CMD position by converting extinction-corrected PHAST colors and magnitudes into a metric ($X_{\rm CMD}$; \citealt{Escala2020b}) \added{analogous to photometric metallicity} that is measured relative to 12 Gyr (F475W, F814W) PARSEC isochrones \citep{Bressan2012,Tang2014,Chen2015} \added{at the adopted M32 distance (see discussion below). In this filter-independent parameterization, $X_{\rm CMD}$ = 0 ($X_{\rm CMD}$ = 1) corresponds to the CMD location of the most metal-poor (metal-rich) isochrone at [Fe/H] = $-2.2$ (+0.5), enabling the identification of blue stars ($X_{\rm CMD} < 0$) that are significantly more likely to belong to the MW foreground population \citep{Gilbert2006}.}
We corrected for the effects of dust extinction caused by the MW foreground by assuming the median value over the low extinction PAndAS footprint ($A_{V, {\rm MW}} = 0.2$; \citealt{McConnachie2018}) based on the dust maps by \citet{Schlegel1998} with modifications by \citet{SchlaflyFinkbeiner2011}, or A$_{\rm F814W} = 0.12$ and A$_{\rm F475W} = 0.38$ when transformed to the appropriate ACS filters \citep{Gregersen2015}. 

Based on homogeneous RR Lyrae-based distances, M32 and M31 are consistent with being located at the same distance ($\mu_{\rm M32} \sim 24.44 \pm 0.06$, or $D_{\rm M31}$ = 20.6$^{+13}_{-21}$; \citealt{Savino2022}), implying that M32 observations are not significantly impacted by dust in M31. Moreover, M32 shows a lack of internal interstellar dust reservoirs (e.g., \citealt{Choi2002,Fiorentino2010,Sarajedini2012}). M31 disk stars present in our dataset may be subject to internal dust extinction \citep{DraineLi2014,Dalcanton2015}, although the stellar properties of RGB stars observed by the SPLASH survey of M31's northeastern disk show little evidence of being affected by internal reddening \citep{Escala2023}.

This parametrization provides a filter-independent means of comparison against the CMD positions of M31/MW SPLASH stars. We conservatively assigned stars with HST colors bluer than the most metal-poor PARSEC isochrone ([Fe/H] = $-2.2$) by an amount greater than the photometric uncertainty to the MW foreground \citep{Gilbert2006,Escala2020b}. Figure~\ref{fig:cmd} shows that this criterion excludes RHeB candidates intentionally placed on the 2022B slitmasks ($ 16.6 \lesssim m_{\rm F814W} \lesssim 20.6$, $2 \lesssim (m_{\rm F475W} - m_{\rm F814W}) \lesssim 3$) that likely correspond to young M31/M32 giant stars with colder kinematics (\citealt{Dorman2015,Quirk2019,Quirk2022}; Grion Filho et al., in preparation). However, as we demonstrate below, the details of the membership determination have little impact on the structure of the resulting velocity distribution.

For some stars in our sample, we were unable to measure \signa\ and/or \sigca\ due to weak absorption, low spectral S/N, or convergence failure in the Gaussian fit of the absorption features \citep{Escala2020b}. Moreover, some stars lack matched PHAST photometry, such that their CMD position could not be used as a diagnostic. In these instances, we used all diagnostic information available to compute a probability of M31/M32 membership, as long as a given star has a successful measurement of \signa, \sigca, or PHAST photometry. 
We excluded \Nnomem\ stars with no available diagnostic measurements from the total sample of \ntargets\ stars with successful velocity measurements from the subsequent analysis.

Figure~\ref{fig:mem} shows the relationship between heliocentric velocity (\vhelio) and \signa\ or \sigca, color-coded by the probability that it belongs to M31 or M32 versus the MW foreground ($p_{\rm M31+M32}$).\footnote{The distribution of EW$_{\rm Na}$ for M31 (the MW) in our model \citep{Escala2020b} has $\langle$EW$_{\rm Na}\rangle$ = 0.52 (3.07) \AA\ and typical errors of $\langle\delta$EW$_{\rm Na}\rangle$ = 1.06 (0.58) \AA.} In general, $p_{\rm M31+M32}$ tends to decrease with increasing \signa\ (higher surface gravity) and decreasing \sigca\ (lower metallicity). Some of the stars with MW-like velocities (\vhelio\ $\gtrsim -100$ \kms) have higher (lower) values of \signa\ (\sigca) and therefore lower M31/M32 probabilities. However, the majority of stars have an overall high probability of being associated with M31 or M32, where the \sigca\ and \signa\ distributions of stars at disk-like (\vhelio\ $\sim$ $-400$ \kms; \citealt{Dorman2012}) and M32-like (\vhelio\ $\sim -200$ \kms) velocities appear indistinguishable. We classified \ngiants\ stars with \added{successful velocity measurements and} a higher likelihood of belonging to M31/M32 than the MW ($p_{\rm M31+M32} > 0.5$) as giant stars.

\begin{figure}
    \centering
    \includegraphics[width=\columnwidth]{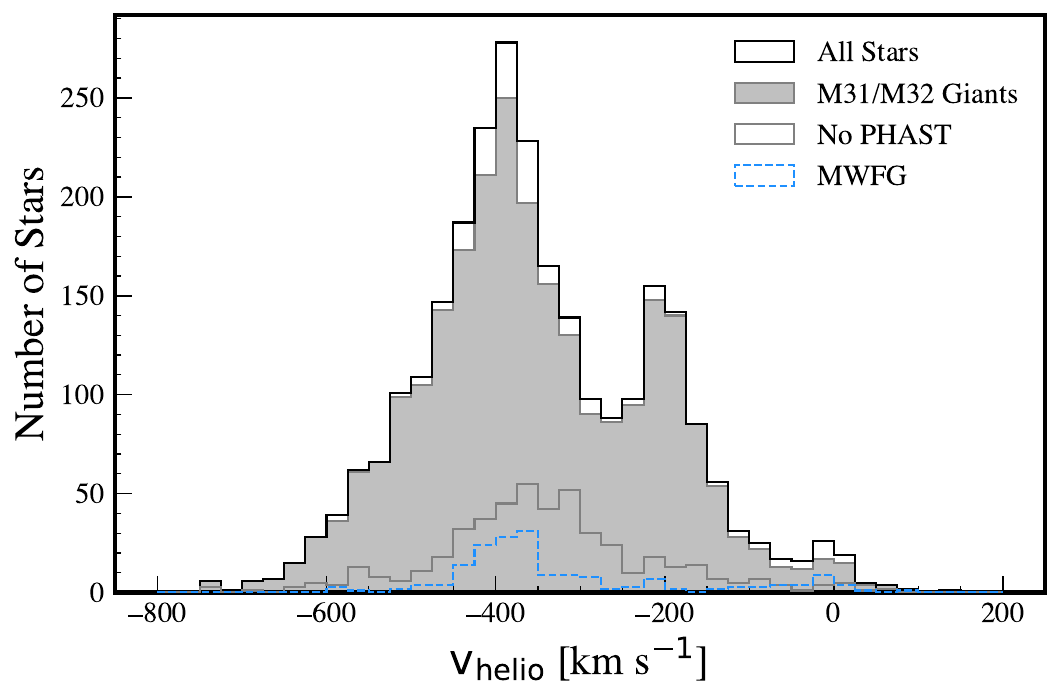}
    \caption{Velocity distributions for stars with successful velocity measurements (black outlined histogram; Section~\ref{sec:obs}) and stars classified as M31/M32 giants (gray filled histogram; Section~\ref{sec:m31bayes}). Stars are present at very negative velocities (\vhelio\ $<$ $-500$ \kms), where the systemic velocity of M31's halo is $-300$ \kms. M31's disk in the vicinity of M32 (Figure~\ref{fig:ring}) exhibits a peak near $-400$ \kms\ and M32 itself is clearly visible as a peak near $-200$ \kms. The membership selection does not significantly alter the velocity distribution, including the tail toward MW-like velocities (near 0 km s$^{-1}$). 
    We also show velocity distributions for stars without a photometric match in PHAST (gray outlined histogram) and stars assigned to the MW foreground (blue dashed histogram). Stars without photometry have velocities similar to M31's disk/halo, whereas MW stars trace the disk (due to RHeB contamination) in addition to MW-like velocities.
    }
    \label{fig:losvd}
\end{figure}

The low contamination fraction of potential MW stars estimated from our sample (\fmwcontamsample) 
is higher than expectations based on the predicted stellar density of the MW foreground relative to M31's stellar disk and M32, but consistent given that our definition includes blue giant stars that are likely at M31 distances. Using the Besan\c{c}on model\footnote{\url{https://model.obs-besancon.fr/modele_home.php}} \citep{Robin2003}, we simulated a population of MW contaminants over the area and magnitude range spanned by the DEIMOS data, assuming all ages, spectral types, and distances out to 150 kpc for
MW stars. We used photometric transformations from \citealt{Sirianni2005} to convert the output BVRI photometry
to the ACS system. We calculated an expected MW contamination fraction of \fmwcontam\ over the portion of the PHAST CMD corresponding to M31/M32 giants, indicating that our membership determination is indeed conservative and that the impact of MW contamination on our sample is likely to be negligible.

We demonstrate the effect of the membership determination on the heliocentric velocity distribution (Section~\ref{sec:obs}) in Figure~\ref{fig:losvd}, where the velocity structure remains largely unchanged between the distributions for all stars with successful velocity measurements and stars classified as M31/M32 giants. The presence of stars at MW-like velocities (near 0 \kms\ in Figure~\ref{fig:losvd}) remains intact following the membership cut, implying that these stars may correspond to a genuine stellar population in the M31 system. 
Figure~\ref{fig:losvd} also shows the velocity distributions for \nnophastgoodposrv\ M31/M32 giant stars without matched PHAST photometry and \nmw\ stars classified as belonging to the MW foreground. The former population exhibits kinematics consistent with being concentrated in M31's stellar disk or halo (where $v_{\rm sys, M31} \sim -300$ \kms), whereas the MW foreground stars are more evenly distributed in velocity. We discuss the velocity distribution structure of M31/M32 giants in more detail in Section~\ref{sec:kinematics}.

\subsection{Stellar Classification}
\label{sec:rgb}

\begin{figure}
    \centering
    \includegraphics[width=\columnwidth]{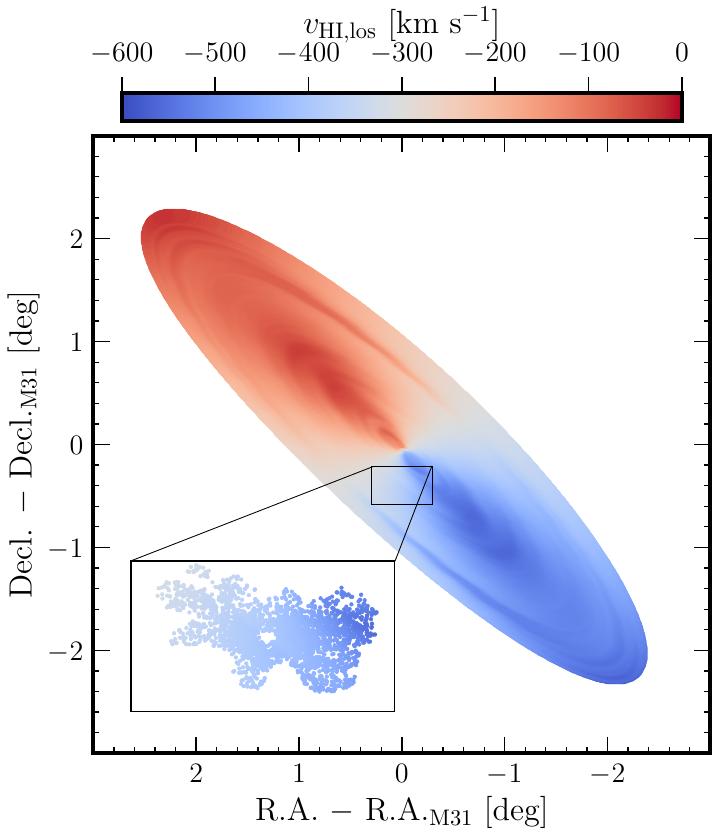}
    \caption{Predicted HI velocity of M31's disk along the line-of-sight (\vHIlos) based on the tilted ring model and HI rotation curve from \citet{Chemin2009} (Section~\ref{sec:ring}). The inset shows the DEIMOS footprint in the vicinity of M32 for stars with successful velocity measurements color-coded by \vHIlos. The median predicted \vHIlos\ over this region is \vlospredmed\ \kms, where M31's disk rotation speed varies from halo-like velocities ($\sim-300$\kms) to velocities characteristic of the southern half of M31's disk ($\sim-500$ \kms).
    }
    \label{fig:ring}
\end{figure}

We categorized M31/M32 giant stars into RGB and AGB stars based on their CMD positions (Figure~\ref{fig:cmd}). These stellar populations trace different mean ages, where RGB (AGB) stars in M32 are $\sim$8.5 (4 $\pm$ 3) Gyr old \citep{Monachesi2011,Monachesi2012}. Moreover, the observed ratio of AGB to RGB stars in this region \citep{Jones2023} indicates a similar average stellar age of 2--4 Gyr for stellar populations in both M32 and M31 \citep{Monachesi2012,Bernard2015a,Williams2015}.

We defined the tip of the RGB (TRGB) using 12 Gyr HST/ACS PARSEC isochrones \citep{Bressan2012} spanning $-2.2 <$ [M/H] $< +0.5$ and a distance modulus of $\mu_{\rm M31} = 24.45\ \pm 0.05$ assuming that M31 and M32 are at the same distance \citep{Savino2022}. Although the light-weighted mean stellar age of M32 is 4.9 Gyr for all stellar populations \citep{Monachesi2012} and equivalently 4 Gyr for RGB stars in M31's disk \citep{Williams2015,Dorman2015}, both systems show evidence of populations with ages at least 10 Gyr old \citep{Fiorentino2012,Williams2017}, motivating the use of the oldest reasonable isochrones to define the AGB, RGB, and MWFG stars to avoid bias against metal-poor populations.
We assigned \nrgb\ stars to the RGB given that they are below the TRGB within the photometric uncertainty: $m_{\rm F814W} + \delta_{\rm F814W} > m_{\rm TRGB}$ (median $\delta_{\rm F814W}$ = \medrederr). We assigned \nagb\ stars above the TRGB but redder than the most metal-poor isochrone to the AGB. We discuss the AGB and RGB velocity distributions in Appendix~\ref{sec:vmodel_rgb}, where we consider the entire M31/M32 giant star velocity distribution in Section~\ref{sec:kinematics}.


\section{Kinematical Modeling of The Stellar Outskirts of M32}
\label{sec:kinematics}

In this section, we model the line-of-sight velocity distribution of the structurally complex region near M32 to isolate the kinematics of M32's resolved stellar outskirts ($R_{\rm M32} \gtrsim$ 1', or $\sim$2\reff). Previous kinematical studies of this region were based on smaller stellar sample sizes, were more limited \added{in} spatial coverage, and relied on dividing the sample into spatial regions to account for the changing dynamical patterns of M31's disk or M32 itself \citep{Dorman2012,Howley2013}. In contrast, we continuously model the kinematics of \ngiants\ M31/M32 giant stars as a single region spanning \rammin--\rammax\ arcmin (or out to $\sim$\rmajeff\reff\ and $\sim$\rmineff\reff\ along M32's major/minor axes) by anchoring the rotation of M31's disk to its HI kinematics, motivated by the formalism by \citet{Gilbert2022} and \citet{Cullinane2023}. 

We model the observed velocity distribution as a combination of stellar populations with kinematics tied to the rotation of M31's gas disk and non-rotating populations with kinematics systematically offset from M31's gas disk.  The former ``rotating'' component captures the velocity distributions of, e.g., thin and thick stellar disks or potentially in-situ stellar halo components formed from the disk (c.f. \citealt{Dorman2013,Gilbert2022,Escala2023,Cullinane2023}), whereas the latter ``offset'' components describe the velocity distributions of dwarf galaxies such as M32, possible coherent tidal debris, or stellar halo components either formed ex-situ from accreted material or in-situ but exhibiting little to no rotation in the disk plane. 

We present the model describing the rotation of M31's stellar disk with respect to the gas in Section~\ref{sec:ring}, where we detail our full ``rotating plus offset'' model 
in Section~\ref{sec:vmodel}.
We explore alternate models including multiple rotating components in Appendix~\ref{sec:altmods}.
We also consider separate ``rotating plus offset'' models with respect to stellar type (and therefore stellar age; Section~\ref{sec:rgb}) in Appendix~\ref{sec:vmodel_rgb}, where the kinematical parameters describing RGB and AGB populations broadly agree. We thus model the M31/M32 giant star velocity distribution as a whole to maximize our sample size and spatial coverage in the vicinity of M32 (Appendix~\ref{sec:vmodel_rgb}).

\subsection{Tilted Ring Model for M31's HI Disk}
\label{sec:ring}

We described M31's HI disk velocity using the tilted ring model of \citet{Chemin2009} (the adopted values in their Table~4), which provides HI rotation speed (\vHIrot), position angle in degrees east of north (PA), and inclination ($i$) as a function of deprojected radius in M31's disk plane (\Rdisk) between 0.4--38 kpc. For each star $j$, we calculated an initial \Rdisk\ estimate based on global values of M31's disk inclination ($i = 77^\circ$) and position angle (PA = 38$^\circ$), then interpolated within tilted ring model based on this initial \Rdisk\ value to assign each star an inclination $i_j$ and position angle PA$_j$. We iteratively revised \Rdisk\ using the updated values of $i_j$ and PA$_j$ until convergence within 0.01$^\circ$ and 0.35$^\circ$ respectively was achieved, following \citet{Gilbert2022} and \citet{Cullinane2023}:
\begin{equation}
    R_{{\rm disk},j} = D_{\rm M31} \sqrt{ \alpha_j^2 + \Bigl[ \frac{\beta_j}{\cos(i_j)} \Bigr]^2 },
\end{equation}
\begin{equation}
    \alpha_j = \eta_j \cos({\rm PA}_j) + \xi_j \sin({\rm PA}_j),
\end{equation}
\begin{equation}
    \beta_j = \xi_j \cos({\rm PA}_j) - \eta_j \sin({\rm PA}_j),
\end{equation}
where ($\xi, \eta$) are the tangent-plane coordinates projected relative to M31's center computed from sky coordinates ($\alpha$, $\delta$) and $D_{\rm M31}$ = 776 kpc \citep{Savino2022}. The line-of-sight component of the HI rotation velocity (\vHIlos) is therefore given by,
\begin{equation}
    v_{\rm HI, los, j} = v_{\rm sys,M31} + V_{\rm HI, rot} \cos (\theta_j) \sin (i_j),
    \label{eq:vHI}
\end{equation}
where $v_{\rm sys,M31} = -300$ \kms\ and $\theta_j$ is the azimuthal angle measured relative to the receding major axis of M31's disk for a given star under the assumption that it is located in the disk plane,
\begin{equation}
    \theta_j = \tan^{-1} \Bigl[ \frac{\beta_j}{\alpha_j \cos (i_j)} \Bigr].
\end{equation}
Figure~\ref{fig:ring} shows \vHIlos\ predicted by the \citet{Chemin2009} tilted ring model, where the rotation of HI gas in M31's disk varies from M31 halo-like velocities ($\sim -300$ \kms) to velocities characteristic of the southern half of M31's disk ($\sim -500$ \kms) over the DEIMOS footprint. The median predicted \vHIlos\ over this region is \vlospredmed\ \kms. Given the substantial variation in line-of-sight velocity for M31's HI disk, we anchored the predicted rotation of M31's \textit{stellar} disk to the gas, accounting for asymmetric drift by parameterizing the stellar disk rotation as a fraction of the HI (\frot),
\begin{equation}
\label{eq:vmod}
    v_{\rm mod, los, j} = v_{\rm sys,M31} + f_{\rm rot} \times V_{\rm HI, rot} \cos (\theta_j) \sin(i_j)
\end{equation}
For each star, we then computed the offset between its observed and predicted line-of-sight velocity (Equation~\ref{eq:vmod}), $v_{\rm offset,j} = v_{\rm mod,los,j} - v_{\rm los,j}$. A stellar population rotating in the plane of M31's HI disk at 
the HI rotation speed will have a \veloffset\ distribution centered at 0 \kms\ with an associated velocity dispersion. At the data location in the southern half of M31's disk, stellar populations leading (lagging) the HI disk rotation will have positive (negative) \veloffset\ values.

\begin{figure}
    \centering
\includegraphics[width=\columnwidth]{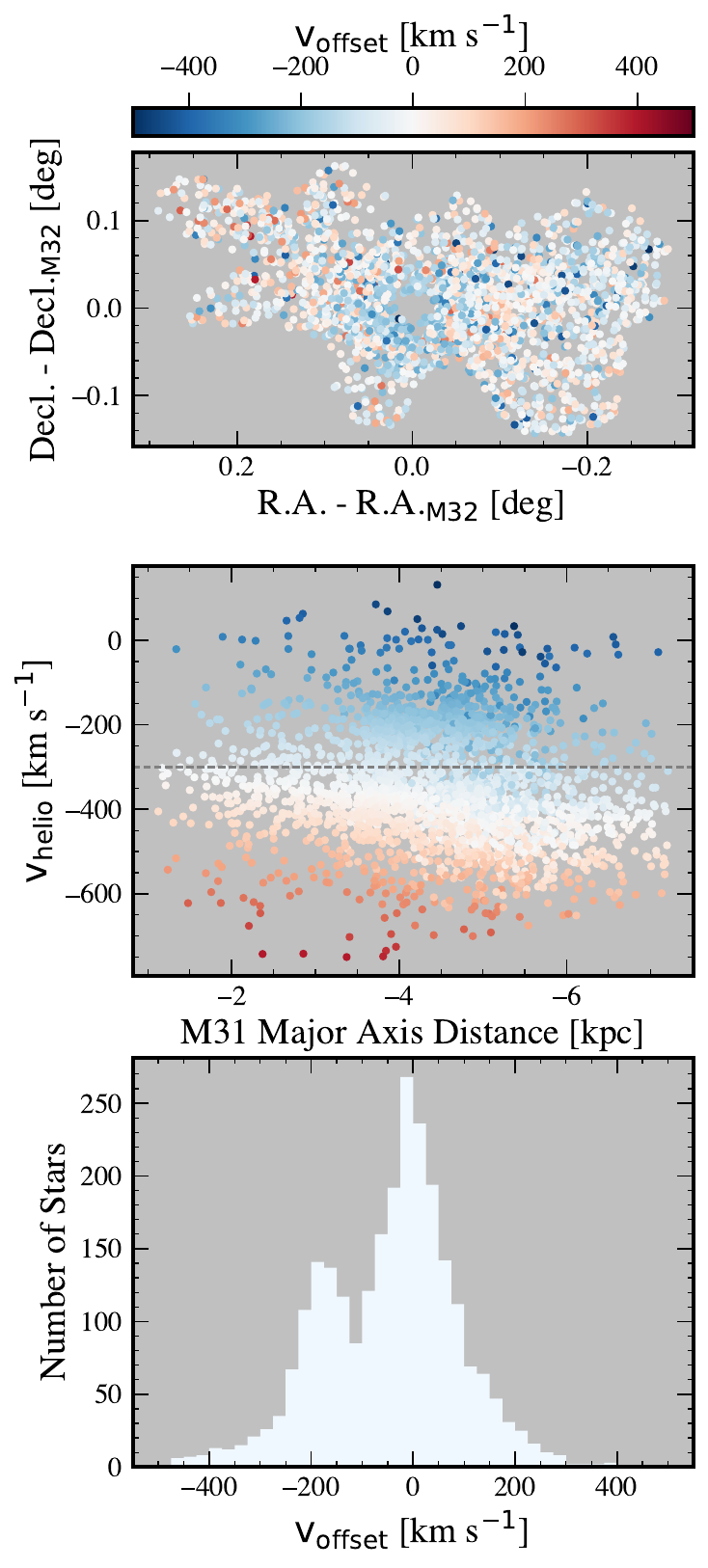}
    \caption{Offset in heliocentric velocity for giant stars near M32 from their predicted M31 HI disk velocity (\veloffset) given the \citet{Chemin2009} tilted ring model (Figure~\ref{fig:ring}; Section~\ref{sec:ring}). (Top) M32-centric location color-coded by \veloffset. (Middle) \vhelio\ vs.\ M31-centric major axis distance color-coded by \veloffset, where $v_{\rm sys,M31} = -300$ \kms (dashed line). The negative sign convention for the M31-centric major axis distance reflects the data location in the southern half of M31's disk. (Bottom) \veloffset\ distribution for M31/M32 giants assuming $f_{\rm rot} = 1$ (light blue filled histogram).
    Stars with \veloffset $\sim$ 0 \kms\ have velocities near M31's predicted disk rotation speed, and stars with \veloffset $>$ 0 \kms\ (\veloffset $<$ 0 \kms) lead (lag) M31's HI disk rotation. Stars near the center of M32 on the sky (and clustered near $R_{\rm maj} \sim -5$ kpc, \vhelio\ $\sim$ $-200$ \kms) tend to lag the rotation speed by $\sim$ 200 \kms.
    }
    \label{fig:voffset}
\end{figure}

The top panel of Figure~\ref{fig:voffset} shows the offset in observed heliocentric velocity for M31/M32 giant stars from the velocity predicted from the HI disk rotation (assuming \frot\ = 1) as a function of sky position. In this disk-centric velocity frame, stars near the center of M32 lag the rotation speed by $\sim$ 200 \kms\ and are distinguishable from disk interlopers with \veloffset\ $\sim$ 0 \kms. Moreover, trends between observed velocity and M31-centric major axis distance ($R_{\rm maj}$) clearly demonstrate multiple stellar populations via M31's rotation curve (\veloffset\ $\sim$ 0 \kms), a cluster of stars corresponding to M32 ($R_{\rm maj}$ $\sim$ $-5$ kpc, \vhelio\ $\sim$ $-200$ \kms), and velocity outliers in the \veloffset\ distribution ($\left| v_{\rm offset} \right| \gtrsim$ 400 \kms).

\subsection{Rotating Plus Offset Model}
\label{sec:vmodel}

Here, we model the velocity distribution of M31's disk in the ``rotating'' velocity frame as a single kinematical component (Eq.~\ref{eq:vmod}) and any additional kinematical components in the ``offset'' or non-rotating heliocentric velocity frame, including an M31 halo-like component. We discuss alternate models that allow for multiple rotating M31 disk components and a rotating M31 halo-like component in Appendix~\ref{sec:altmods}. The likelihood of our fiducial model is given by a sum of Gaussian components,
\begin{equation}
\begin{split}
    \mathcal{L} = \sum_j & \biggl[ \Bigl( 1 - \sum_N f_N \Bigr) \times \mathcal{N}\bigr(v_{\rm offset,j} (f_{\rm rot})|0,\sigma_{\rm disk}\bigl) \\
    &+ \sum_N f_N \times \mathcal{N}(v_{\rm helio,j} | \mu_N, \sigma_N) \biggr],
\label{eq:func}
\end{split}
\end{equation}
where $N$ is the number of offset model components, $f$, $\mu$, and $\sigma$ are the fractional contribution, mean velocity, and velocity dispersion of a given offset component, and \sigmadisk\ is the velocity dispersion of M31's stellar disk. The probability that an individual star $j$ is associated with a given offset component assuming the best-fit model parameters $\Theta$ is therefore,
\begin{equation}
    p_{j,N} = \frac{f_N \times \mathcal{N} (v_{{\rm helio},j} | \mu_N, \sigma_N )}{ \mathcal{L}(\Theta)_j },
\label{eq:likelihood}
\end{equation}
and analogously for the rotating component in the \veloffset\ velocity frame. We motivate the adopted number of model components and discuss their potential physical nature in Section~\ref{sec:ncomp} and describe the assumed priors on these components in Section~\ref{sec:prior}. We present the best-fit kinematical parameters of the rotating plus offset model (Equation~\ref{eq:func}) in Section~\ref{sec:results}. We construct visualizations of the model (Figures~\ref{fig:vmodel},~\ref{fig:vmodel_unif}) by transforming the rotating component into the heliocentric velocity frame following \citet{Gilbert2022} (their Appendix~B).

We acknowledge various caveats to our model, including the assumption that rotating model components do so in the disk plane, such that the model cannot adequately capture rotation in other planes. Owing to the tight coupling between the stellar and gas disks, the model also assumes that M31's giant stars follow the same PA and inclination trends with radius as the HI (Section~\ref{sec:ring}) and have circular orbits. However, these limitations affecting the modeling of M31's stellar disk are minimized by the small area covered by the data (Figure~\ref{fig:obs}). 

\subsubsection{Number of Model Components}
\label{sec:ncomp}

The structure of the velocity distributions in Figures~\ref{fig:losvd} and~\ref{fig:voffset} provides clear visual evidence for at least two kinematical components corresponding to M31's disk and M32. However, we anticipate that M31's stellar halo and bulge should also contribute to the stellar populations in this region (Section~\ref{sec:2022B}), whereas the non-negligible numbers of stars at the tails of the distribution in excess of expectations based on M31's halo velocity dispersion (Section~\ref{sec:prior}; Table~\ref{tab:prior}) suggests the possible presence of additional kinematical components. 

We estimated the marginal likelihood $\mathcal{Z}$ of various $N$-component models given the data using the dynamic nested sampling implementation {\sc DYNESTY} \citep{Skilling2006,Higson2019,Speagle2020},
\begin{equation}
    \mathcal{Z} = \int \mathcal{L} (\Theta) \pi ( \Theta ) d \Theta,
    \label{eq:dlogz}
\end{equation}
where $\pi(\Theta)$ is the prior associated with the model parameters. We then compared the logarithm of the marginal likelihood ($\log_{10} \mathcal{Z}$) to evaluate the statistical evidence for increasing the number of components. 

For this exercise, we assumed weakly informative uniform priors on all model parameters to remain relatively agnostic about the physical interpretation of each component. The first component, which is tied to M31's HI gas rotation, corresponds to M31's stellar disk, whereas all subsequent components are ``offset'' from the gaseous disk. We provide the details of the nested sampling implementation in Appendix~\ref{sec:appendix}, which includes Table~\ref{tab:uniform} summarizing the comparison between various $N$-component models and Figure~\ref{fig:vmodel_unif} showing the best-fit models compared to the observed velocity distribution.

With the weakly informative uniform priors, we found evidence for M31's stellar disk and \added{four} offset components, or \added{five} total kinematical components. While the first offset component has high velocity dispersion (\sigmav\ $\sim$ \added{100} \kms) similar to M31's stellar halo \citep{Dorman2012,Gilbert2018}, it has a mean velocity (\muv\ $\sim$ $-500$ \kms) significantly offset from M31's systemic velocity ($v_{\rm sys,{\rm M31}} \sim -300$ \kms), at odds with expectations for M31's halo. \added{The second offset component has M31 disk-like mean velocity, but low velocity dispersion (\muv\ $\sim-390$ \kms, \sigmav\ $\sim30$ \kms), which could correspond to a secondary disk component (Appendix~\ref{sec:altmods}).} 
The \added{third} offset component has a mean velocity close to M32's systemic velocity ($-200$ \kms) and is kinematically cold similar to a dwarf galaxy.
The \added{fourth} offset component ($\mu_v \sim$ \added{$-60$} \kms, $\sigma_v \sim$ \added{80} \kms) accounts for stars at the positive tail of the velocity distribution. In this model, the stellar disk slightly lags the gaseous disk (\frot\ $\sim$ \added{0.85}) with a velocity dispersion (\sigmadisk\ $\sim$ \added{80} \kms) 
\added{in accordance with expectations for this region} \citep{Dorman2012}.

The observed velocity distribution that we aim to model represents a structurally complex region on the sky, where the relative fraction of different model components changes significantly as a function of position. The combination of the complexity of the observed velocity distribution and the model assumption of weakly informative priors may therefore result in the attribution of stars belonging to one kinematical component to another. We thus explored the effect of separately modeling the kinematics of spatial regions near and far from M32 on the nature of the recovered model components, assuming the same weakly informative uniform priors as used in modeling the full dataset. We assigned M31/M32 giant stars with surface-brightness based M32 probability \psb\ $>$ 0.01 to an ``M32 Region'' and stars with \psb\ $<$ 0.01 to a ``Disk Region'' (Section~\ref{sec:2022B}).

\begin{figure*}
    \centering
    \includegraphics[width=\textwidth]{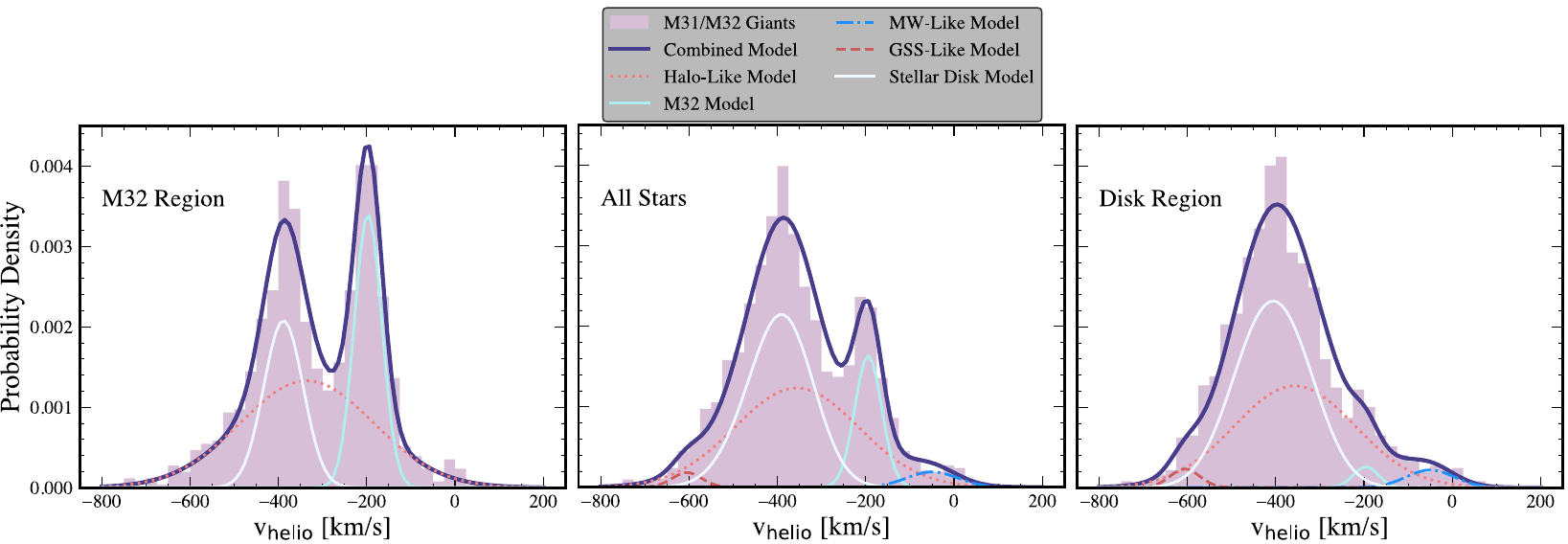}
    \caption{Observed velocity distributions for M31/M32 giant stars compared to best-fit velocity models assuming normal priors (Table~\ref{tab:prior}), where the models consist of a component rotating in the velocity frame relative to M31's disk (\veloffset; Figure~\ref{fig:voffset}) at some fraction of the predicted rotation speed (``Stellar Disk Model'') and multiple components offset from the rotating component in the heliocentric velocity frame (Section~\ref{sec:vmodel}; Table~\ref{tab:offset_model_norm}). We show best fit models for giant stars near M32 (left panel; \psb\ $>$ 0.01; Figure~\ref{fig:pm32sb}), away from M32 (right panel; \psb\ $<$ 0.01), and all giant stars (middle panel). We find evidence for four total ``offset'' components corresponding to M32, two populations with MW-like and GSS-like mean velocities respectively (although not necessarily associated with these structures), and an M31 halo-like population with high velocity dispersion. Stars with M32-like velocities (\vhelio\ $\sim-200$\kms) are still present in the ``Disk Region'', demonstrating that M32 velocity structure extends beyond the region predicted by surface brightness alone.
    }
    \label{fig:vmodel}
\end{figure*}

Figure~\ref{fig:vmodel} 
shows the velocity distributions for these spatial regions, where the M32 region has two clear peaks corresponding to M31's stellar disk and M32. Stars with M32-like velocities are still present in the disk region, demonstrating that M32's velocity structure is likely to extend beyond the spatial region predicted by surface brightness alone.
In the disk region, we identified M31's stellar disk ($\sigma_v \sim$ \added{70} \kms), a kinematically hot component significantly offset from M31's systemic velocity (\muv\ $\sim$ \added{$-430$} \kms), and a kinematically \added{colder} component at the positive tail of the velocity distribution that is distinct from M32. \added{We also recover a component at M32-like velocities (\muv\ $\sim-180$ \kms) with intermediate velocity dispersion (\sigmav\ $\sim$ 50 \kms).}
This is qualitatively similar to the model components identified from the full dataset, \added{excepting the secondary disk-like component suggested from all giant stars.}
In the M32 region, we found evidence for M31's stellar disk ($\sigma_v \sim$ \added{40} \kms), a kinematically hot component with mean velocity close to M31's systemic velocity, and a kinematically cold component with mean velocity close to M32's systemic velocity (Table~\ref{tab:uniform}; Figure~\ref{fig:vmodel_unif}). The presence of a single kinematically hot component, as found in the M32 region, better conforms to expectations for M31's stellar halo than the two 
\added{kinematical components with high and intermediate velocity dispersion at the tails of the velocity distribution} suggested by the disk region.

In choosing our fiducial model of one rotating component and four offset components, we combine the results of the above exercise with our existing knowledge of the properties of M31's structural components, specifically its stellar halo and the presence of GSS-related tidal debris, derived from analyses covering much larger regions of M31 \citep{Gilbert2009,Dorman2012,Gilbert2018}.
We thus interpret the two kinematically hotter offset components preferred under the assumption of weakly informative priors as the likely superposition of three total components: an M31 halo-like component centered near M31's systemic velocity plus two kinematically colder components at the positive (MW-like) and negative (GSS-like) velocity tails of the distribution (Section~\ref{sec:prior}). \added{Given that the secondary M31 disk-like component is only identified in the full dataset, and not the disk region, we assume a single rotating component tied to HI rotation curve for simplicity (Appendix~\ref{sec:altmods}).}
Moreover, the fact that \sigmadisk\ decreases in accordance with literature expectations (Table~\ref{tab:uniform}) when modeling the M32 and disk regions separately implies that weakly informative priors limit the ability of the model to disentangle multiple distinct kinematical components. The results of this exercise based on weakly informative priors therefore indicates the need for more informative priors (Section~\ref{sec:prior}).

Given the above arguments, we adopt a fiducial model consisting of five components---a rotating M31-disk like component and four offset non-rotating components---assuming informative priors on the model parameters for each component based on literature results (Section~\ref{sec:prior}). However, in Section~\ref{sec:results}, we also explore the results of assuming fewer model components while retaining informative priors.

\subsubsection{Normal Priors}
\label{sec:prior}

\begin{table}
    \centering
    \label{tab:prior}
    \caption{Normal Prior Parameters for ``Offset'' Model}
    \begin{threeparttable}
    \begin{tabular*}{\columnwidth}{@{\extracolsep{\fill}}l|cccc}
    \hline\hline
    Comp.\@ & Parameter &  \multicolumn{1}{p{1.2cm}}{\centering Mean [km s$^{-1}$]} &  \multicolumn{1}{p{1.2cm}}{\centering Std.\@ [km s$^{-1}$]} & Ref. \\ \hline
    Disk & $\sigma_{\rm disk}$ & 68.0 & 26.6 & 1 \\ \hline
    Halo-Like & $f_v$ & 0.45 & 0.05 & 1\\
    & $\mu_v$ & $-$337.0 & 9.3 & 1,5 \\
    & $\sigma_v$ & 126.8 & 9.3 & \\
    \hline
    M32 & $\mu_v$ &  $-197.8$ & 4.4 & 4,5 \\
    & $\sigma_v$ & 28.2 & 4.4 & \\ \hline
    MW-Like & $\mu_v$ & $-55.9$ & 8.8 & 2,5 \\
    & $\sigma_v$ & 44.2 & 8.8 & \\ \hline
    GSS-Like & $\mu_v$ & $-618.0$ & 8.8 & 3,6 \\
    & $\sigma_v$ & 30.0 & 8.8 & \\
    \hline
    \end{tabular*}
\end{threeparttable}
\begin{tablenotes}[flushleft]
\item \footnotesize References. \textemdash\ (1) \citet{Dorman2012}, (2) \citet{Gilbert2012}, (3) \citet{Fardal2012},  (4) H13, (5) \citet{Gilbert2018}, (6) \citet{Escala2022}. 
The priors for all other values of $f_v$ (M32, MW-like, and GSS-like components), as well as for $f_{\rm rot}$ for M31's disk, are uniform over the ranges [0,1] ($f_v$) and [0,1.5] ($f_{\rm rot}$).
\end{tablenotes}
\end{table}

We assumed normal priors on the mean velocity (\muv) and velocity dispersion (\sigmav) parameters fitted for each ``rotating plus offset'' velocity model (Table~\ref{tab:offset_model_norm}), where we summarize the adopted priors for each physically motivated kinematical component in Table~\ref{tab:prior}. As discussed in Section~\ref{sec:ncomp} and Appendix~\ref{sec:appendix}, we 
assumed five total kinematical components corresponding to M31's stellar disk, a high velocity dispersion component similar to M31's stellar halo, M32, and two components at the negative and positive velocity tails of the distribution with mean velocities similar to expectations for GSS-related tidal debris in this region and the MW foreground population (although not necessarily associated with either population).

For the stellar disk of M31, we assumed uniform priors on \frot\ between 0 and 1.5  to allow the model to explore rotation speeds from those approaching zero to speeds similar to or slightly faster than predicted from the HI. We computed the Gaussian prior on \sigmadisk\ from an average of the disk dispersion parameters derived by \citet{Dorman2012} in their ``SSW'' region corresponding to the H13 spectroscopic sample, where we excluded the anomalously low dispersion value fitted in the angular subregion containing M32 (their Table~4). 

For the halo-like component, we calculated the priors on \muv\ and \sigmav\ from a simple average of two sources: (1) the kinematical M31 spheroid parameters from \citet{Dorman2012} for the entire SSW region corrected for tidal debris (their Table~3), and (2) the kinematical stellar halo parameters for projected M31-centric radii between 8--14 kpc from \citet{Gilbert2018}, which we transformed from the galactocentric to the heliocentric frame based on the mean R.A.\ and Decl.\ of our sample \citep{vanderMarel2008,vanderMarel2012}. This assumes that M31's stellar halo exhibits similar dynamical behavior at our data location ($R_{\rm proj,M31}$ $\sim$ \rprojmin--\rprojmax\ kpc) compared to 8--14 kpc. 

We also implemented a normal prior on the fractional contribution ($f_v$) of the halo-like component in the 5D model (Section~\ref{sec:ncomp}) based on the simple mean of $f_v$ values in the SSW region of \citet{Dorman2012} (again excluding the subregion containing M32). Without this constraint on the prior, the fitting procedure prefers a model where $f_{v, {\rm halo}}$ $\sim$ \fvhalonoprior\ (in contrast to the $f_v$ values suggested by all other models in Table~\ref{tab:offset_model_norm}, as well as literature results based on the H13\ subsample) in favor of increasing the disk dispersion to $\sigma_{v, {\rm disk}}$ $\sim$ \sigmadisknofhprior\ \kms\ and decreasing the disk rotation speed to \frot\ $\sim$ \frotnofhprior. 

The data prefer a substantial M31 halo-like contribution in the vicinity of M32, despite surface-brightness based predictions from a structural decomposition in M31's northeast implying an M31 halo-like contribution of \percenthaloliketot\ (for the combination of M31's stellar halo and bulge over the survey area; Section~\ref{sec:2022B}, \added{\citealt{Dorman2013}}). \added{However, M31's global surface brightness profile is known to substantially underpredict M31's halo fraction in the vicinity of M32 compared to observational constraints from velocity distribution modeling, which find consistently high M31 halo fractions ($\gtrsim$40--50\%) in this region despite differences in methodology and spectroscopic samples (\citealt{Dorman2012,Howley2013}; this work). This partially results from the fact that surface brightness is \textit{not} equivalent to RGB star counts, where the luminosity of M31's young stellar disk per RGB star is high relative to M31's old stellar halo. Moreover, the substantial M31 halo fraction near M32 could also result from a high velocity dispersion M31 disk that is difficult to distinguish from a kinematically hot halo-like component \citep{Dorman2013,Escala2023}, and possibly the presence of unknown tidal material with high velocity dispersion.
}

\begin{table*}
    \centering
    \caption{Kinematical Parameters for Rotating Plus Offset Model -- Normal Priors}
    \label{tab:offset_model_norm}
    \begin{threeparttable}
    \begin{tabular*}{\textwidth}{@{\extracolsep{\fill}}l|cc|ccccc}
        \hline\hline
        Model &  $f_{\rm rot}$ & $\sigma_{\rm disk}$ [km s$^{-1}$] & Comp. & $f_v$ & $\mu_v$ [km s$^{-1}$] & $\sigma_v$ [km s$^{-1}$] \\ \hline
         \multicolumn{1}{c}{} & \multicolumn{2}{c|}{Disk} &  \multicolumn{3}{c}{Offset Components} &  \\ \hline
         
        \multicolumn{7}{c}{Preferred Model, All Stars}\\ \hline
        5D$^\ast$ & $0.99^{+0.01}_{-0.03}$ & $62.2^{+2.7}_{-6.3}$ & Halo-Like &  $0.43^{+0.02}_{-0.05}$ & $-356.1^{+4.0}_{-9.2}$ & $139.7^{+2.6}_{-6.7}$ \\
        & & & M32 &  $0.12^{+0.01}_{-0.01}$ & $-194.2^{+1.0}_{-2.4}$ & $30.4^{+1.2}_{-2.4}$  \\
        & & & MW-Like & $0.03^{+0.00}_{-0.01}$ & $-49.4^{+3.2}_{-8.4}$ & $57.6^{+2.5}_{-6.3}$  \\
        & & & GSS-Like & $0.02^{+0.00}_{-0.01}$ & $-604.6^{+4.5}_{-10.9}$ & $33.5^{+3.3}_{-6.5}$  \\ \hline
        
        \multicolumn{7}{c}{Alternative Model, All Stars}\\ \hline
        3D & $1.01^{+0.01}_{-0.03}$ & $63.3^{+2.5}_{-5.2}$ & Halo-Like & $0.49^{+0.02}_{-0.04}$ & $-340.1^{+2.6}_{-5.8}$ & $156.5^{+1.7}_{-3.7}$  \\
        & & & M32 & $0.10^{+0.01}_{-0.01}$ & $-195.0^{+1.1}_{-2.6}$ & $27.9^{+1.2}_{-2.5}$  \\ \hline
        
        \multicolumn{7}{c}{Disk Region, $p_{\rm M32, SB} < 0.01$} \\ \hline 
        5D & $0.98^{+0.02}_{-0.04}$ & $67.7^{+2.1}_{-4.9}$ & Halo-Like & $0.43^{+0.02}_{-0.04}$ & $-356.9^{+3.8}_{-8.9}$ & $136.7^{+2.7}_{-6.3}$  \\ 
        & & & M32 & $0.02^{+0.00}_{-0.01}$ & $-195.8^{+2.0}_{-4.9}$ & $28.3^{+2.0}_{-4.5}$ \\
        & & & MW-Like & $0.03^{+0.00}_{-0.01}$ & $-50.0^{+3.7}_{-8.8}$ & $54.3^{+3.0}_{-6.9}$  \\
        & & & GSS-Like & $0.02^{+0.00}_{-0.01}$ & $-605.8^{+4.0}_{-10.3}$ & $33.8^{+3.4}_{-8.2}$ \\ \hline
        
        \multicolumn{7}{c}{M32 Region, $p_{\rm M32, SB} > 0.01$} \\ \hline   
        3D & $1.08^{+0.03}_{-0.07}$ & $43.0^{+2.8}_{-5.7}$ & Halo-Like & $0.50^{+0.02}_{-0.04}$ & $-335.5^{+3.0}_{-6.9}$ & $150.9^{+2.1}_{-4.9}$  \\
        & & & M32 & $0.27^{+0.01}_{-0.02}$ & $-197.0^{+1.1}_{-2.5}$ & $31.1^{+1.0}_{-2.4}$ & \\ \hline
        
\end{tabular*}
\begin{tablenotes}[flushleft]
    \item$^{\ast}$ The fiducial 5D model (Section~\ref{sec:results}) for all M31/M32 giant stars is adopted for the analysis of M32's kinematics in Section~\ref{sec:m32}.
\end{tablenotes}
\end{threeparttable}
\end{table*}

The priors for M32 are based on a simple average of the values from \citet{Dorman2012} (their Table~4) and H13. For the MW-like component at the positive tail of the velocity distribution, we adopted values based on the mean and standard deviation of the line-of-sight velocity distribution for securely identified MW stars in the SPLASH survey of M31's southeastern minor axis \citep{Gilbert2012,Gilbert2018}. We relied on this empirical distribution rather than model predictions for the MW foreground at the data location (e.g., \citealt{Robin2003}; Section~\ref{sec:m31bayes}) because the observed distribution of MW stars is known to differ owing to the selection function of our surveys \citep{Gilbert2018}. The standard deviations on the priors for the MW-like (and GSS-like) component(s) are equivalent to twice that of the M32 priors (4.4 \kms), which is approximately the median velocity uncertainty of our sample (\vtotmed\ \kms).

We based the priors for the ``GSS-like'' component on N-body model predictions for the velocity distribution of satellite tidal debris from the GSS merger event at the data location \citep{Escala2022}. In this model, the GSS and its associated tidal shells form from a progenitor with stellar mass $M_{\rm sat} = 2.2 \times 10^9\ M_\odot$ that completely disrupted 0.8 Gyr ago, where this model broadly reproduces the observed properties of M31's major system of tidal features \citep{Fardal2007,Fardal2012,Fardal2013,Gilbert2007,Escala2022}. In Appendix~\ref{sec:gss}, we show the predicted debris pattern in projected phase space compared to the observed distribution of M31/M32 giants. Although the model predicts M31 to dominate the stars in this region, GSS-related tidal debris should be most detectable near $-600$ \kms, where stars with such negative velocities are present in our dataset. Moreover, the predicted debris also extends to positive MW-like velocities (near 0 \kms).

We isolated the dominant tidal feature in the model using a two-component 
variational Bayesian Gaussian mixture \citep{scikitlearn}, where the predicted stream has \muv\ $\sim$ $-618$ \kms\ and \sigmav\ $\sim$ 30 \kms\ (Table~\ref{tab:prior}). Despite being limited to a single model in a specific merger scenario, this theoretical constraint is useful for evaluating the statistical significance of the identification of a stellar population potentially originating from the GSS progenitor. 

\subsubsection{Resulting Velocity Models}
\label{sec:results}

We sampled the posterior distribution of the model parameters using the dynamic nested sampling implementation DYNESTY \citep{Speagle2020} in order to maximize the logarithm of the model likelihood (Equation~\ref{eq:likelihood}). We present the best-fit kinematical parameters for ``rotating plus offset'' models assuming 5 total components across the entire spectroscopic sample (Section~\ref{sec:ncomp}) in Table~\ref{tab:offset_model_norm},  where parameter values are based on the 16$^{\rm th}$, 50$^{\rm th}$, and 84$^{\rm th}$ percentiles of the marginalized posterior probability distribution. The offset model components consist of M32, a high-dispersion M31 halo-like population, and two components at GSS-like and MW-like velocities (Section~\ref{sec:prior}). 

Figure~\ref{fig:vmodel} shows the best-fit models for giant stars near M32 (\psb\ $>$ 0.01), away from M32 (\psb\ $<$ 0.01), and all giant stars covering the region spanned by our entire dataset. For all M31/M32 giant stars, M32 contributes $\sim$\fracdwarf\ of the stars in the sample with $\mu_{v, {\rm M32}}$ = \muvdwarf\ \kms\ and $\sigma_{v,{\rm M32}}$ = \sigmavdwarf\ \kms, consistent with the results of restricting the model to the region near M32. \added{We also detect a small contribution from M32 at the $\sim$2\% level in the region away from M32 and dominated by the disk of M31.}

Across all spatial regions, we found consistent values for the M31 halo-like component of \muv\ $\sim$ $-340$ \kms\ and $\sigma_v$ $\sim$ 145 \kms, which contributes $\sim$43\% of the stars when considering a more globally representative sample that includes stars away from the M32 region. 
M31's stellar disk rotates with speeds similar to the gas (\frot\ $\sim$ 1) and a dispersion of $\sim$60 \kms, though we find a faster rotation speed and smaller dispersion for the portion of the disk in the localized region near M32 (similar to the results of \citealt{Dorman2012}).

We also found evidence for small contributions (at the $\sim$2--3\% level) from populations with GSS-like and MW-like velocities when considering spatial regions containing stars with low prior probability of belonging to M32. We assessed whether including the GSS-like and MW-like model components indeed provides a better description of the data by comparing the fiducial 5D model to a 3D model including only M31's stellar disk, an M31 halo-like component, and M32 (Table~\ref{tab:offset_model_norm}). The 3D model provides a worse description of the data ($\Delta \log \mathcal{Z} \sim$ \dlogzthreevsfive). 
Regardless of whether we included GSS-like or MW-like components, the M32 kinematical parameters are consistent between the 3D and 5D models in Table~\ref{tab:offset_model_norm}.

Despite the presence of stars with velocities near $-600$ \kms\ and 0 \kms\ (Figure~\ref{fig:vmodel}), we do not detect GSS-like or MW-like populations at a statistically significant level in the M32 region, likely owing to the restricted coverage of this subsample.  We confirmed that including a GSS-like or MW-like component is not preferred over a model excluding these components in the M32 region ($\Delta \log \mathcal{Z}$ $< 0$). 
In Section~\ref{sec:m32}, we base the analysis of M32's kinematics  on the fiducial 5D model for all M31/M32 giant stars, which provides the best statistical description of the data over the largest area.

\section{Kinematical Properties of Stellar Populations in M32}
\label{sec:m32}

Here, we explore M32's internal dynamical properties 
in the context of our spatially continuous model (Section~\ref{sec:kinematics}) for the line-of-sight velocity distribution on this region of the sky.  In Section~\ref{sec:losvd}, we search for differences in the structure of M32's probabilistically extracted velocity distribution between its ``inner'' and ``outer'' regions, separated by the radius at which M32's isophotal contours in its surface brightness profile are observed to twist along its major axis ($R_{\rm iso} \sim 150'' \sim 0.56$ kpc; Figure~\ref{fig:pm32sb}; C02). 
We model the rotational behavior of M32's inner and outer regions in Section~\ref{sec:rot} and construct on-sky maps of M32's predicted velocity structure enabled by our modeling approach in Section~\ref{sec:maps}.  

Throughout this section, we use the surface-brightness based M32 probability ($p_{\rm M32,SB}$; Section~\ref{sec:2022B}) only as a guide on where to expect M32 stars compared to predictions from kinematics. As noted by H13, the idealized 2D surface brightness model may not accurately capture non-uniform deviations in the surface brightness of M31's disk, or may reflect an excessively steep surface brightness profile for M32. Moreover, the global surface brightness model for M31 is based on constraints from the northeast \citep{Dorman2013}, and may not be representative of the region of the sky near M32.

\subsection{M32's Line-of-Sight Velocity Distribution}
\label{sec:losvd}

\begin{figure*}
    \centering
     \includegraphics[width=0.8\textwidth]{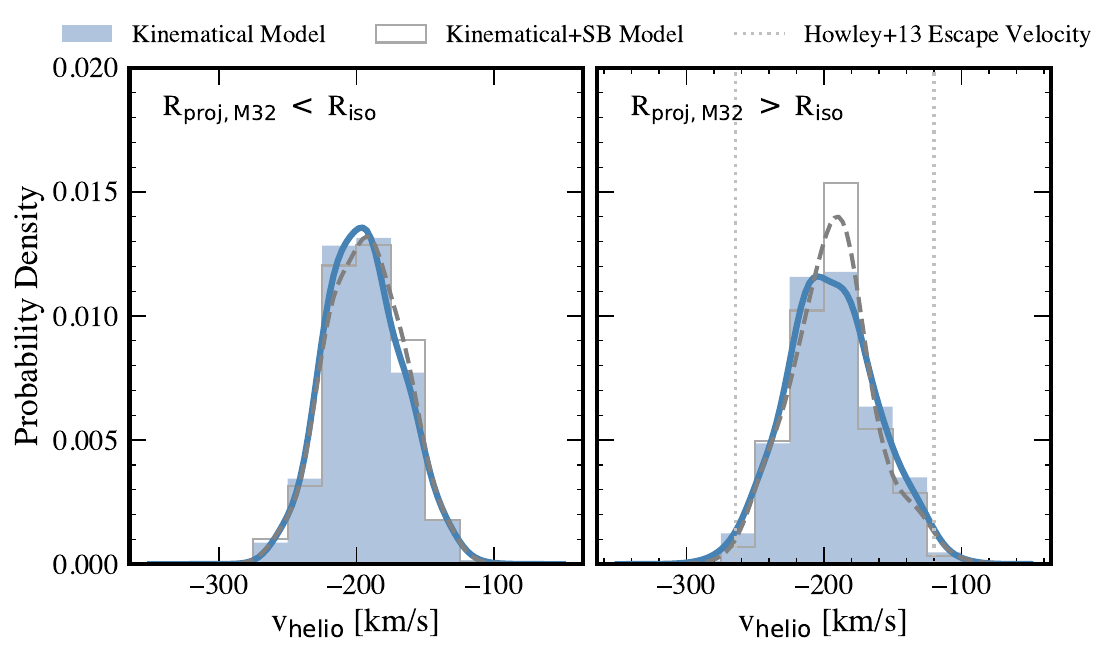}
    \caption{Velocity distributions for M32's inner (left panel) and outer regions (right panel), separated by the isophotal twisting radius ($R_{\rm iso} \sim$ \risoas). We show histograms of stellar velocities weighted by M32 probability, based on kinematics alone ($p_{\rm M32, vel}$, blue histograms) and including surface brightness information ($p_{\rm M32, vel} \times p_{\rm M32,SB}$, grey histograms), and also weighted by the inverse square of the velocity measurement uncertainty. We also show corresponding Gaussian kernel density estimates (blue solid lines and grey dashed lines, respectively). For reference, we include M32's escape velocity at 200'' in the equatorial plane calculated from a three-integral model by H13 ($v_{\rm M32}$ $\pm$ 72 \kms) in the right panel. Beyond the twisting radius, M32 
    has a 
    velocity distribution with heavier outliers \added{and moderately higher dispersion 
    compared to the} 
    distribution within $R_{\rm iso}$. 
    }
    \label{fig:m32losvd}
\end{figure*}

Prior studies with more limited sample size and coverage, as well as distinct modeling approaches, found that M32's kinematics appeared symmetric and without clear evidence of sharp gradients across the region defined by M32's isophotal elongation (H13). We thus revisit M32's kinematical properties with respect to $R_{\rm iso}$ in light of our improved sample size and coverage and updated statistical methodology. 

Figure~\ref{fig:m32losvd} shows the velocity distributions for M32's inner and outer regions, constructed by weighting the observed line-of-sight velocities by their position and velocity based probabilities of belonging to M32 (Equation~\ref{eq:likelihood}; Section~\ref{sec:results}). We show velocity distributions for M32 extracted using probabilities from solely the kinematical model ($p_{\rm M32, vel}$) and the combination of the kinematical plus surface-brightness models ($p_{\rm M32, vel} \times p_{\rm M32, SB}$).  Regardless of whether surface-brightness information is included,
M32's predicted velocity structure does not substantially change for M32's inner regions. In contrast, the model including surface-brightness information predicts that M32's outer velocity distribution 
is more centrally concentrated compared to the case without surface-brightness information. As shown in Section~\ref{sec:ncomp}, photometry alone predicts fewer stars at larger M32-centric radii compared to kinematics (see also H13), thereby decreasing the probability of belonging to M32 beyond $R_{\rm iso}$. Given that our goal is to constrain M32's kinematical structure in its outskirts, we therefore focus on predictions based solely on $p_{\rm M32, vel}$.

 Figure~\ref{fig:m32losvd} further demonstrates that M32's velocity structure changes between its inner and outer regions. Inside the isophotal twisting radius, M32 has a velocity distribution characterized by negative kurtosis, or a centrally concentrated flat peak with few outliers, that results from M32's ordered rotation.  Beyond this radius, M32's velocity distribution 
 \added{has larger dispersion and more} positive kurtosis \added{(consistent with zero)}, indicating
 heavier outliers. We quantified the change in M32's velocity structure by computing the moments of the distributions based on 1000 Monte-Carlo bootstrap trials. We perturbed the sample velocities assuming Gaussian velocity uncertainties, then re-computed the velocity distribution, weighting each velocity measurement by $p_{\rm M32,vel}$, and measured the moments of the resulting bootstrapped distribution. We computed the median values and the uncertainties based on the 16$^{\rm th}$, 50$^{\rm th}$, and 84$^{\rm th}$ percentiles of this distribution. 
 For $R_{\rm proj,M32} < R_{\rm iso}$, we obtained $\mu_v$ = \meaninner\ \kms,  $\sigma_v$ = \stdinner\ \kms, $\gamma_v$ = \gammainner, and $\kappa_v$ = \kappainner\ for the mean, standard deviation, skewness, and excess kurtosis respectively. For $R_{\rm proj,M32} > R_{\rm iso}$, we found $\mu_v$ = \meanouter\ \kms,  $\sigma_v$ = \stdouter\ \kms, $\gamma_v$ = \gammaouter, and $\kappa_v$ = \kappaouter.\added{\footnote{The mean velocity $\mu_v$ decreases slightly with increasing M32-centric radius, which may be a consequence of M32's rotation over the data footprint (Section~\ref{sec:rot}), minor asymmetric velocity structure in M32's outskirts, or contamination from M31 stars at more negative velocities and larger M32-centric distances. The latter explanation is unlikely given that the difference in $\mu_v$ across $R_{\rm iso}$ becomes more significant when including \psb\ to produce a more centrally concentrated M32 sample. Moreover, the rotational models in Section~\ref{sec:rot} predict that M32's mean velocity should become more \textit{positive} with increasing M32-centric distance given the spatial distribution of our sample.}}
 \added{Thus, both $\sigma_v$ and $\kappa_v$ differ across $R_{\rm iso}$ by $\sim$2.2$\sigma$ ($\sim$1.1$\sigma$ when including \psb) and $\sim$1.7$\sigma$ ($\gtrsim$3$\sigma$ with \psb) respectively.}
 When considered with M32's other internal kinematical properties (Section~\ref{sec:rot}), we interpret these changes in kinematical structure as evidence in favor of tidal distortions affecting M32 (Section~\ref{sec:discuss}).

 For reference, we also show M32's escape velocity at $R_{\rm proj, M32}$ = 200''  in Figure~\ref{fig:m32losvd}, which was calculated from a three-integral Schwarzschild orbit model for M32's mass distribution fitted to a combination of integrated-light and resolved stellar kinematics by H13 ($v_{\rm M32}$ $\pm$ 72 \kms, where we adopt $v_{\rm M32}$ from Table~\ref{tab:offset_model_norm}).  The probability density of the predicted velocity distribution approaches zero near this escape velocity, although some velocity structure remains present beyond this boundary.  We estimated the predicted fraction of stars beyond this velocity using Gaussian kernel density estimates of M32's line-of-sight velocity distribution (as in Figure~\ref{fig:m32losvd}), which are weighted by the inverse square of the velocity measurement uncertainty and the probability of belonging to M32 ($p_{\rm M32, vel}$). 
 
 We found that approximately  \fracunbound\% of stars  (\fracunboundsb\% when incorporating $p_{\rm M32,SB}$) 
associated with M32 in our model beyond 200'' may have velocities beyond the escape velocity, in qualitative agreement with the findings of H13 that stars potentially associated with M32 are present beyond this limit.  
 If these stars beyond the isophotal twisting radius and the escape velocity are bound to M32, H13 interpreted them as possible evidence of a dark matter halo, non-equilibrium tidal distortions in M32's outskirts, or M31 substructure. However, they found the explanation of a dark matter halo less likely given M32's asymmetric velocity distribution, where the excess of fast-moving stars at large radii were preferentially detected toward M32's positive velocity tail ($\sim$\fracunboundpos\% out of $\sim$\fracunboundhow\% of stars beyond 200'' in our model when assuming $v_{\rm M32}$ = $-201$ \kms\ as in H13). We further discuss these hypotheses for explaining M32's observed kinematics in Section~\ref{sec:discuss}.
 
 \subsection{Rotation in M32}
 \label{sec:rot}

 \begin{figure*}
    \centering
    \includegraphics[width=\textwidth]
    {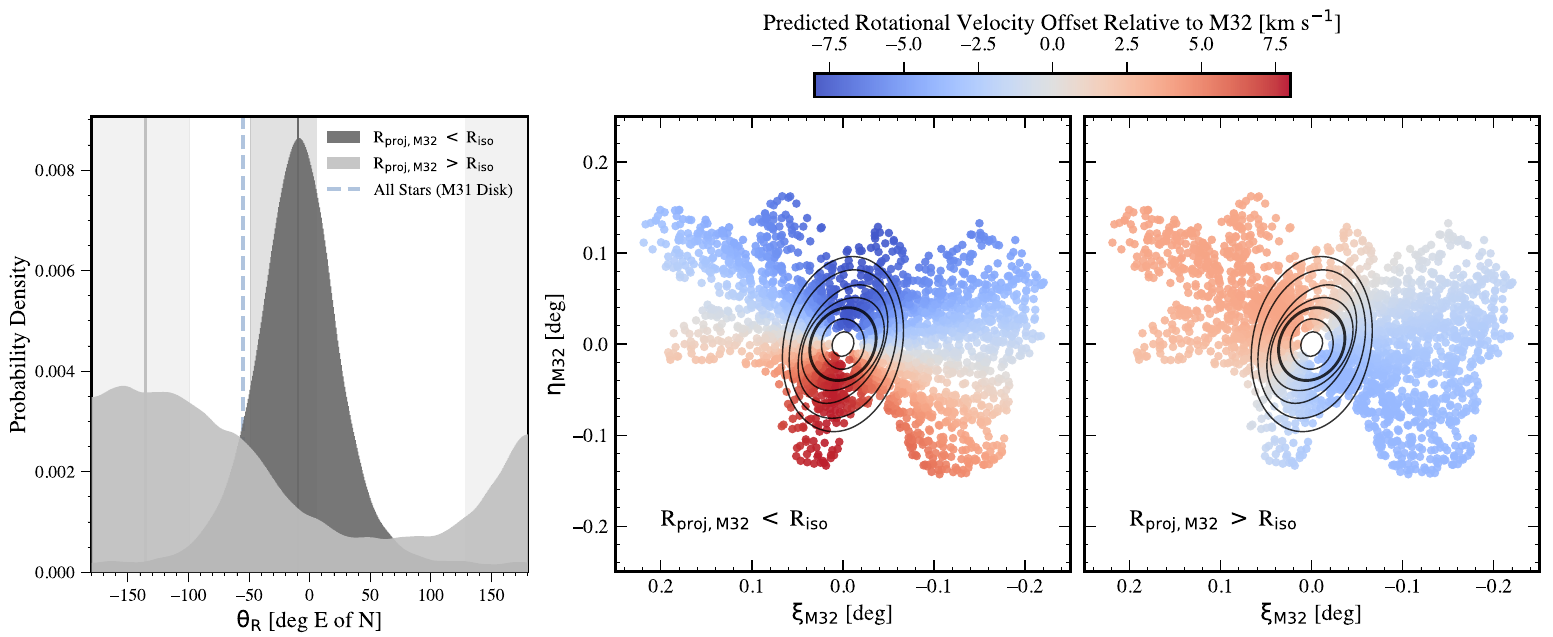}
    \caption{
    Rotational model parameters describing M32's line-of-sight velocity distribution (Equation~\ref{eq:vrot}; Section~\ref{sec:rot}), derived by weighting by the probability of belonging to M32 ($p_{\rm M32,vel}$). (Left panel) Marginalized posterior probability distributions for the rotational direction ($\theta_R$) within (dark grey) and outside of (light grey) the isophotal twisting radius ($R_{\rm iso}$), with vertical lines showing median values and shaded regions showing 1$\sigma$ confidence intervals. 
    We also show $\theta_R$ for all M31/M32 giant stars (i.~e., unweighted by $p_{\rm M32,vel}$), which corresponds to a population dominated by M31's disk (solid blue line).
    (Right panels) 
    M32-centric location of each star in our dataset color-coded by its predicted 
    velocity offset, due to M32's rotation, from the mean velocity of M32 (Table~\ref{tab:offset_model_norm}).
    We also show the I-band isophotes from C02 for reference, with $R_{\rm iso}$ 
    bolded (black ellipses).
    The orientation of the rotation (or linear velocity gradient, \added{a model which more precisely constrains $\theta_{\rm outer}$}; see Section~\ref{sec:rot}) shifts direction from the region inside 
    and outside 
    $R_{\rm iso}$; within $R_{\rm iso}$, the rotation is approximately aligned along M32's major axis ($\theta_{R,{\rm inner}} \sim$ \thetarinner; receding from NNW to SSE, as expected), whereas outside $R_{\rm iso}$ the change in velocity is aligned roughly along M32's minor axis ($\theta_{R, {\rm outer}} \sim$ \thetarouter; receding from WSW to ENE). This direction is distinct from that of the rotation of M31's disk over the data footprint ($\theta_{R, {\rm disk}} \sim$ \thetardisk; receding from NW to SE; Figure~\ref{fig:ring}).
    }
    \label{fig:rot}
\end{figure*}

 M32 is known to possess a steep rotational velocity gradient at its center driven by a central black hole (\mbh; e.g., \citealt{Bender1996,vanderMarel1997a,vanderMarel1998,Joseph2001,Verolme2002,vandenBosch2010}). M32's rotational velocity falls from maximum value of $v_{\rm max} \sim$ \vmaxrot\ \kms\ at a radius of $\sim$1'' (4 pc) along its major axis before flattening at $\sim$10 \kms\ beyond $\sim$100'' or 0.4 kpc, whereas no clear line-of-sight velocity trends have been detected along its minor axis (H13). We therefore modeled M32's rotation based on our computed M32 probabilities ($p_{\rm M32,vel}$; Equation~\ref{eq:likelihood}), separately considering populations located inside and outside of M32's isophotal twisting radius. The likelihood function describing the observed line-of-sight velocity due to M32's azimuthally uniform rotation is,
\begin{equation}
\begin{split}
    \mathcal{L}_j &= \mathcal{N} \bigl( v_{{\rm helio},j} | \mu_{\rm M32} - k_{\rm R} \cos(\theta_{R} - \theta_j), \sigma_{\rm M32} \bigr),\\
    \log \mathcal{L} &= \sum_j p_{{\rm M32,vel},j} \times \log \mathcal{L}_j
    \label{eq:vrot}
\end{split}
\end{equation}
where $k_R$ is the rotation magnitude, $\theta_R$ is the position angle representing the rotation direction in degrees east of north, and $\mu_{\rm M32}$ and $\sigma_{\rm M32}$ are the adopted parameters describing M32's velocity distribution from Table~\ref{tab:offset_model_norm}.

We followed the procedure detailed in Section~\ref{sec:results} to sample the associated posterior probability distribution, assuming a normal prior of \kvpr\ on $k_R$, based on the major-axis rotational profile of H13 at large radii (their Figure 10) and our typical velocity uncertainty (Table~\ref{tab:prior}), and a uniform prior bounded by [0, 2$\pi$] on $\theta_R$ (which we converted to degrees after sampling). 
We found best-fit values of $k_R$ = \krinner\ \kms\ and $\theta_R$ = \thetarinnerval\ deg E of N 
for the rotational velocity magnitude and direction respectively within $R_{\rm iso}$ ($k_R$ = \krouter\ \kms\ and $\theta_R$ = \thetarouterval\ deg E of N \added{for a weak rotational signature} outside of $R_{\rm iso}$).

The rotational velocity magnitude for both M32's inner and outer region is consistent within the uncertainties with previous results from binned spatial regions in M32's outskirts along its major axis from H13. Although the errors on the 
rotational direction are large, the marginalized posterior probability distribution \added{for M32's inner region} has a shape similar to a skewed Gaussian that is clearly peaked about the median value (Figure~\ref{fig:rot}). Moreover, the best-fit value of $\theta_R$ within $R_{\rm iso}$ is close to M32's major axis direction (roughly $-22$ deg E of N; C02) as expected, despite the fact that we 
used uniform priors on $\theta_R$. This supports the interpretation that the true value of the 
rotation direction is likely accurately, albeit imprecisely, recovered by our model.

Figure~\ref{fig:rot} shows the velocity offset, due to M32's rotation, from M32's mean velocity (Equation~\ref{eq:vrot}) predicted over the data footprint both inside and outside of $R_{\rm iso}$, color-coded by the value assigned to each star assuming it belongs to M32. The orientation of M32's rotation between its inner and outer regions is predicted to change: the line-of-sight velocity shifts from increasing from NNW to SSE along M32's major axis within $R_{\rm iso}$, to increasing from WSW to ENE along M32's minor axis outside of $R_{\rm iso}$. Despite the large uncertainties in $\theta_R$, the values for $\theta_{R, {\rm inner}}$ and $\theta_{R, {\rm outer}}$ are non-overlapping within the 1$\sigma$ confidence intervals (Figure~\ref{fig:rot}). This supports the statistical significance of a change in the rotational direction when considered 
with the fact that the model is able to accurately recover M32's major-axis rotation within $R_{\rm iso}$.

As noted in the case of Milky Way satellite dwarf galaxies that are likely tidally disrupted (e.g., \citealt{Ji2020}), it is not necessarily straightforward to disentangle observational signatures of rotation from tidally induced linear velocity gradients owing to the similar functional forms when modeling line-of-sight velocity distributions \citep{Walker2016,Caldwell2017}.  This raises the question of whether the change in $\theta_R$ between M32's inner and outer regions indeed reflects a change in M32's rotation direction, or a transition to a regime characterized by tidal distortions, as opposed to ordered rotation, that coincides with M32's isophotal elongation. To demonstrate this, we modified Equation~\ref{eq:vrot} to depend on the projected distance from the center of M32 ($R_{\rm M32}$), 
\begin{equation}
    \mathcal{N} \bigl( v_{\rm helio},j | \mu_{\rm M32} - k_{\rm V} R_{\rm M32} \cos(\theta_{V} - \theta_j), \sigma_{\rm M32} \bigr),
    \label{eq:grad}
\end{equation}
where $k_V$ and $\theta_V$ correspond to the magnitude and direction of the 
linear velocity gradient respectively. 

Assuming that M32 rotates at $v_{\rm rot} \sim$ 10 \kms\ over a region encompassed by $|R_{\rm M32}| <$ 1 kpc (approximately twice $R_{\rm iso}$) similarly yields an expected prior on $k_V$ of $\mathcal{N}$(10,4.4) for the velocity gradient magnitude. From sampling the posterior distribution,  we obtain a weak gradient of $k_V$ = \kvouter\ \kms\ kpc$^{-1}$ and $\theta_V$ = \thetavouter\ deg E of N for $R_{\rm proj, M32} > R_{\rm iso}$ (\added{$k_V$ = \kvinner\ \kms\ kpc$^{-1}$ and $\theta_V$ = \thetavinner\ deg E of N for $R_{\rm proj, M32} < R_{\rm iso}$}). 
These model parameters produce a similar predicted line-of-sight velocity pattern on the sky compared to the rotational model, \added{albeit with more precise constraints on the gradient direction for $R_{\rm proj, M32}$ $>$ $R_{\rm iso}$,} supporting the notion that M32's detected rotation beyond the isophotal twisting radius can be reasonably modeled as a weak linear velocity gradient.

The shift in M32's rotational direction with respect to $R_{\rm iso}$ may be a kinematical signature of tidal distortion (Section~\ref{sec:discuss}), but could also originate from contamination by rotating stellar populations distinct from M32 that become more dominant at large M32-centric radii. For example, M31's disk exhibits significant rotation over the data footprint (Figure~\ref{fig:ring}; Section~\ref{sec:ring}), and M31's inner stellar halo is also known to rotate at $V_{\rm rot, halo} =  52.6 \pm 6.8$ \kms\ for $ R_{\rm proj,M31}$ $>$ 5 kpc in approximately the same direction as M31's disk (\citealt{Dorman2012}; Appendix~\ref{sec:altmods}). 

\begin{figure*}
    \centering
    \includegraphics[width=0.9\textwidth]{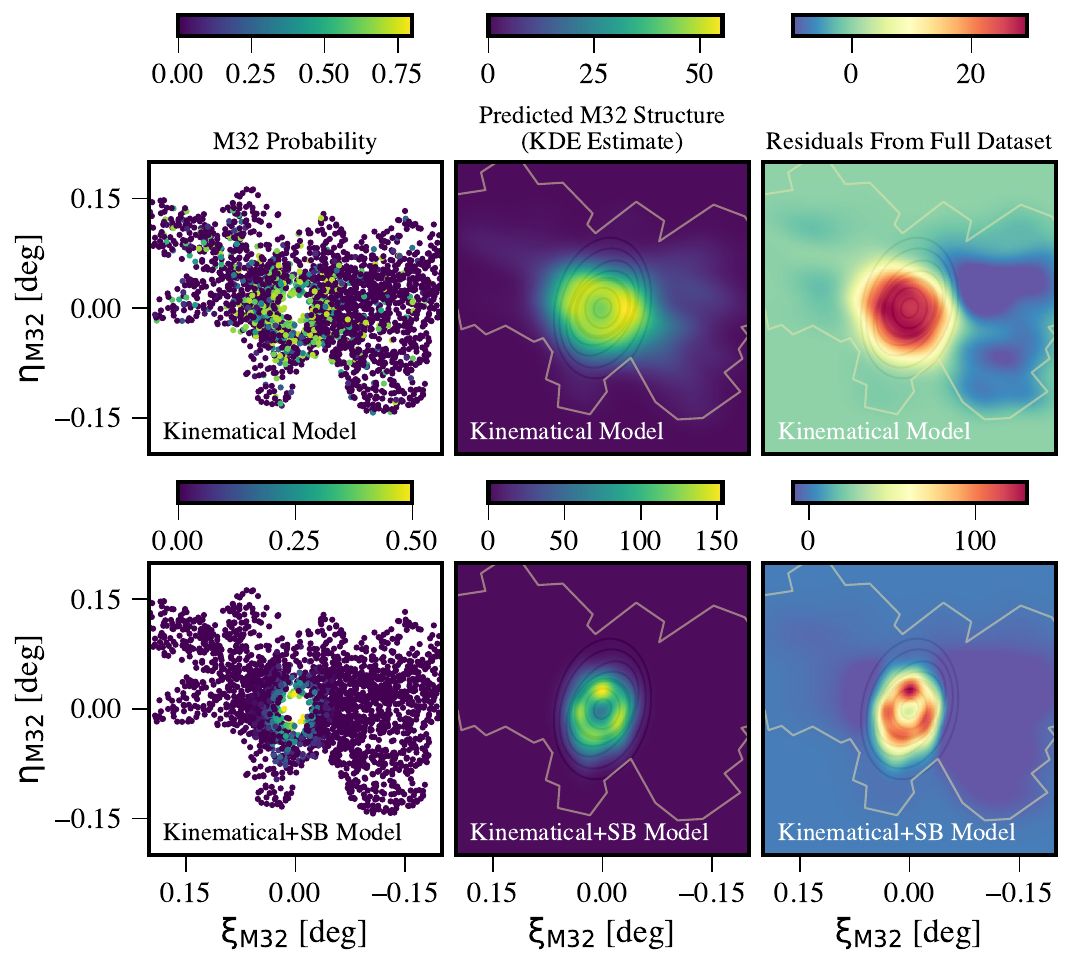}
    \caption{(Left panels) Location of M31/M32 giant stars in our dataset on the sky, color-coded by probability of M32 association based on their spatial positions and velocities predicted from kinematical modeling (Section~\ref{sec:vmodel}). We show predictions from $p_{\rm M32,vel}$ (top panels; Equation~\ref{eq:likelihood}) and $p_{\rm M32, vel}$ $\times$ $p_{\rm M32, SB}$ (bottom panels; Section~\ref{sec:2022B}). The net effect of incorporating $p_{\rm M32,SB}$ is decreasing $p_{\rm M32}$ overall, particularly at large M32-centric radii.
    (Middle panels) Relative 2D probability density of M32 on the sky 
    constructed from a Gaussian kernel density estimate weighted according to 
    probability of belonging to M32 and the inverse square of the velocity measurement uncertainty. I-band isophotes for M32 are overlaid, with $R_{\rm iso}$ bolded (black ellipses; C02). We also show the data footprint from Figure~\ref{fig:obs} for reference (yellow outlines).
    (Right panels) Residuals between maps of M32 and similar maps created using all M31/M32 giant stars in the dataset (i.\@e.\@ unweighted by $p_{\rm M32}$). 
    The residual maps based on kinematical models alone (top panels) show that M32 is more likely to be identified toward the southeast (away from M31), with tentative evidence of structure beyond the outermost isophote. The maps incorporating surface-brightness information (bottom panels) show density enhancements near the isophotal twisting radius ($R_{\rm iso}$ $\sim$ 5\reff\ $\sim$ \risoas\ or \risokpc\ kpc). 
    }
    \label{fig:maps}
\end{figure*}

We therefore tested whether $\theta_{R, {\rm outer}}$ is distinct from the sense of M31's disk/halo rotation ($\theta_{R, {\rm disk}}$) by sampling the posterior distribution of Equation~\ref{eq:vrot} \textit{without} weighting by $p_{\rm M32, vel}$, where our sample is dominated by stars in M31's disk and halo-like populations ($\sim$\fracdiskhaloinsample\ of the sample; Table~\ref{tab:offset_model_norm}). Assuming uniform priors on both parameters, we measured $k_{R, {\rm disk}}$ = \krdisk\ \kms,  $\theta_{R, {\rm disk}}$ = \thetardiskval\ deg E of N, corresponding to a strong disk rotational signature over the data footprint. This orientation is similar to that of M32 within $R_{\rm iso}$, where both $\theta_{R, {\rm inner}}$ and $\theta_{R, {\rm disk}}$ are receding towards the northwest, whereas the direction for M32 beyond $R_{\rm iso}$ ($\theta_{R, {\rm outer}}$) recedes to the southwest. 
We thus conclude that the shift in M32's rotational direction with $R_{\rm iso}$ is unlikely to be driven by contamination from M31's disk or halo, given that the distinct orientation of M32's rotation beyond the isophotal twisting radius ($\theta_{R, {\rm outer}}$ = \thetarouterval\ deg E of N) cannot be produced by a summation of M32's inner rotation and M31 contamination.

\subsection{Predicted Structural Maps of M32}
\label{sec:maps}

We constructed 2D maps of M32's structure on the sky predicted by our spatially continuous ``rotating plus offset'' model from Section~\ref{sec:vmodel}. The upper left panel of Figure~\ref{fig:maps} shows the location of all M31/M32 giants in our dataset color-coded by the likelihood-based probability of belonging to M32 computed from the spatial positions and line-of-sight velocities of individual stars ($p_{\rm M32, vel}$; Equation~\ref{eq:likelihood}). Despite the fact that the kinematical model does not incorporate information on M32's sky location (Section~\ref{sec:vmodel}), stars near the center of M32 have the highest probability of being members of the dwarf galaxy, although some stars with M32-like velocities are present at large M32-centric radii.

We approximated the probability density function (PDF) of M32's spatial distribution using a Gaussian kernel density estimate weighted according to $p_{\rm M32, vel}$ and the inverse square of the velocity measurement uncertainty. The resulting map in the upper middle panel of Figure~\ref{fig:maps} shows that the kinematical model predicts faint stellar structures associated with M32 beyond its outermost I-band isophote (C02), which appear as elongations NE (SW) in M32's north (south).\footnote{Although no data is present in M32's central regions (left panels of Figure~\ref{fig:maps}), the PDF and associated residual maps (center and right panels of Figure~\ref{fig:maps}) are non-zero, albeit still low density, at their centers as a consequence of the Gaussian kernel.
}

These potential structures could result from contamination by stars in M31's disk and/or halo with M32-like velocities that were incorrectly attributed to M32 by the model. However, given that the model explicitly accounts for the line-of-sight velocity variation of M31's disk, we do not expect M31's disk to be a significant source of contamination, particularly west of M32 where M31's disk has more negative predicted velocities relative to M32 (Figure~\ref{fig:ring}). Moreover, the shapes of the putative structures to the NE and SW of M32 beyond the outermost isophote could reflect the data footprint or selection function (Section~\ref{sec:obs}) instead of M32's underlying stellar density distribution. 

To evaluate the 
robustness 
of the features compared to the data footprint, we constructed maps of the residuals between PDFs for M32 and the full dataset of M31/M32 giant stars (i.e., between PDFs weighted and unweighted by $p_{\rm M32,vel}$). The upper right panel of Figure~\ref{fig:maps} shows that M32 is preferentially detected toward the southeast (in the direction away from M31; Figure~\ref{fig:obs}) and largely within $R_{\rm iso}$. The model also preferentially removes stars to the west of M32, where the velocity difference between M31's disk and M32 is the largest ($\mu_{v, {\rm disk}} \sim -500$ \kms\ vs.\ $\mu_{v, {\rm M32}} \sim -200$ \kms; Figure~\ref{fig:ring}). Figure~\ref{fig:maps} also illustrates that elongations in M32's predicted stellar structure beyond the outermost isophote remain present in the NE--SW direction. 
Interestingly, this direction roughly corresponds to M32's minor axis and therefore the change in orientation of M32's rotational velocity signature beyond $R_{\rm iso}$ (Section~\ref{sec:rot}, see also~\ref{sec:discuss}).

The bottom panels of Figure~\ref{fig:maps} are the same as the top panels, except using M32 probabilities including surface-brightness information ($p_{\rm M32, vel} \times p_{\rm M32, SB}$). The net effect of incorporating $p_{\rm M32,SB}$ is to decrease the total M32 probability, particularly at large M32-centric radii, thereby increasing PDF amplitudes near M32's center. However, as discussed in Section~\ref{sec:losvd}, the predicted M32 fractions from 2D surface-brightness profiles tend to be systematically lower at a given radius than those from the line-of-sight velocity distribution 
(H13). In the maps highlighting the predicted structure in M32's inner regions, the most notable features to emerge are the density enhancements located along M32's major axis and near $R_{\rm iso}$. These density enhancements do not correspond to highly sampled regions in the data (e.g., the HST/ACS PM field; Figure~\ref{fig:obs}) and neither do they appear in 2D PDFs that we constructed from weighting by $p_{\rm M32, SB}$ alone as a test of whether these features are intrinsic to the surface brightness profile and the data footprint. 

\section{Summary \& Discussion}
\label{sec:discuss}

Based on a sample of \ngiants\ individual giant stars with radial velocities from Keck/DEIMOS spectroscopy along the line-of-sight to M32, we have modeled M32's kinematics between \rammin--\rammax\ arcmin from its center. Motivated by the methodology presented in \citealt{Gilbert2022} and \citealt{Cullinane2023}, we have developed a framework to model the full line-of-sight velocity distribution as the combination of M31's rotating stellar disk, which is anchored to its HI rotation curve \citep{Chemin2009}, and non-rotating components systematically offset from M31's disk, which correspond to M32, a high-velocity dispersion population representing M31's stellar halo, and any additional substructure that may be present in M31. In contrast to the only prior resolved kinematical study of M32 (H13), this novel approach enables spatially continuous modeling of the full survey area in a Bayesian context to directly address the complication of M32's projected location at the edge of M31's disk. 
Given our expanded sample size ($\sim$3.5$\times$ larger usable unique target velocities) and spatial coverage (out to $\sim$\rmajeff$r_{\rm eff}$ and $\sim$\rmineff$r_{\rm eff}$ along M32's major and minor axes 
vs.\ $\sim$8$r_{\rm eff}$ for H13), this also allows us to revisit the kinematical evidence---or lack thereof---for signatures of tidal distortion in M32.

We analyzed the kinematics of M32 within and outside of the radius at which its I-band isophotes begin to twist and elongate ($R_{\rm iso} \sim$ \risoas $\sim$ \risokpc\ kpc; C02) to search for any corresponding changes in M32's velocity structure. \added{We find minimal evidence for a change in the mean velocity} inside and outside $R_{\rm iso}$, in agreement with H13. 
\added{However,} we find \added{suggestions of an increase in the velocity dispersion and} a change from negative ($\kappa_v \sim -0.3$) to \added{approximately zero} excess kurtosis. This indicates that M32's inner velocity distribution 
has relatively few outliers, 
which may be due to its ordered rotation,
\added{compared to the heavier outliers in M32's outer velocity distribution} (Figure~\ref{fig:m32losvd}). These outliers may exceed M32's escape velocity ($v_{\rm M32} \pm 72$ \kms; H13). Indeed, we find evidence for stars associated with M32 beyond its tidal radius ($R_{\rm tidal} \sim $ 1.2 kpc; C02), where stars at M32-like velocities are present even when surface brightness information alone predicts the contribution from M32 should be negligible (Figure~\ref{fig:vmodel}).  If these stars are bound to M32, we interpret them as evidence for either a dark matter halo or non-equilibrium tidal distortions in M32's outskirts (see H13 and Section~\ref{sec:losvd}). We deem the third possible explanation of these stars belonging to unknown M31 substructure to be less likely, given that we have explicitly included components to account for this possibility in our kinematical model (Section~\ref{sec:ncomp}).

In addition to investigating M32's observed line-of-sight velocity distribution, we modeled M32's rotation with respect to $R_{\rm iso}$, with the rotational axis as a free parameter (Section~\ref{sec:rot}). We recover a weak rotational signature along M32's major axis within $R_{\rm iso}$ ($k_R \sim 7$ \kms, $\theta_R \sim$ \thetarinner\ deg E of N), in accordance with expectations from H13. Interestingly, we find evidence of a shift in M32's rotational direction beyond $R_{\rm iso}$ to approximately along M32's minor axis ($k_R \sim 5$ \kms, $\theta_R \sim$ \thetarouter\ deg E of N; Figure~\ref{fig:rot}), acknowledging that line-of-sight rotational signatures can be equivalently modeled as linear velocity gradients. In maps of M32's structure on the sky predicted by our kinematical model (Figure~\ref{fig:maps}), we similarly find hints of structural elongations in M32's minor-axis direction (Section~\ref{sec:maps}). Based on the expected rotational pattern of M31's stellar disk 
(Figure~\ref{fig:ring}; Section~\ref{sec:ring}),  it is unlikely that these features are caused by M31 contamination. 

Taken together with the change in M32's velocity distribution with respect to $R_{\rm iso}$ and the presence of stars with M32-like velocities at large M32-centric radii (Section~\ref{sec:losvd}),  we argue that this presents compelling evidence that M32's stellar outskirts have been tidally distorted by interactions with M31.\footnote{Alternately, M32's outskirts may have contributions from a ``fossil'' disk component as purported by \citealt{Graham2002}, which could be supported by the change in M32's rotational pattern (Section~\ref{sec:rot}). In this case, the progenitor would have still experienced tidal stripping by M31. Interactions with other M31 dwarf galaxies may also contribute to tidal distortions in M32.
} 
Although prior observational constraints from H13 and C02 provided minimal apparent support for tidal distortions, this conclusion was largely limited to M32's central $\sim$1 kpc ($\lesssim8r_{\rm eff}$ or $\lesssim1.5 R_{\rm iso}$). In contrast, our detection of distinct M32 kinematics beyond $R_{\rm iso}$ is based on a larger radial range ($\gtrsim3 R_{\rm M32}$, or $\lesssim6 R_{\rm iso}$).
If M32's progenitor had an intrinsically compact and dense core, it is reasonable to expect little tidal evolution in its central regions (see discussion by \citealt{DSouzaBell2018}). 
Yet the existence of tidal distortions in the outskirts of M32 does not provide decisive evidence in favor of M32 being the remnant of a tidally stripped disky progenitor (e.g., \citealt{Bekki2001}), given that modest amounts of  stripping are still expected in scenarios where M32's progenitor is a dwarf galaxy of similar present-day stellar mass (\citealt{Dierickx2014}, see also \citealt{Du2019}).

Detailed kinematics for cEs near massive hosts beyond the Local Group could place M32 in context toward informing its most likely formation pathway. However, owing to the lack of nearby cEs that can be resolved into individual stars, kinematical information for cEs in the local universe is limited to coarse rotational and velocity dispersion profiles from integrated light spectroscopy (e.g., \citealt{FerreMateu2021}). Although not classified as a cE, the nearby dwarf elliptical satellite galaxy of M31, NGC 205, provides a comparison system to M32 owing to its similarly elliptical morphology and observed twisting of isophotes in its surface brightness profile (C02). Based on the abrupt turnover in the major axis velocity profile at the isophotal twisting radius \citep{Geha2006}, NGC 205 has likely experienced a strong tidal interaction with M31 in the past  \citep{Howley2008,Widmark2025}. In particular, the orientation of NGC 205's major-axis rotation flips beyond this radius, where the stellar motion in NGC 205's outskirts implies it is on a prograde encounter with M31 \citep{Geha2006}. In comparison to NGC 205, H13 concluded that M32's kinematics appeared symmetric across M32's major axis, without clear gradients across M32's region of isophotal elongation. However, the shift in M32's rotational direction beyond its isophotal twisting radius (Section~\ref{sec:rot}) could imply that similar physical mechanisms are at play in M32 and NGC 205. If these mechanisms are tidal in origin, they may depend on the intrinsic rotational properties and orbit of each galaxy.

The current sample of existing models for tidal encounters between M32 and M31 provides limited insight in terms of expected kinematical effects on M32. C02, who focused on purely photometric indicators, argued that the detection of twisted isophotes in M32 indicated an inclined orbital plane relative to M32's direction of motion ($i \lesssim 70^\circ$). Furthermore, they inferred that the location of the breaks in the surface brightness profile implies a highly eccentric orbit with M32 currently near apocenter ($\pm$10 kpc relative to M31) based on comparisons to simulations of satellites \citep{Johnston2002}. If these inferred orbital properties are accurate, the orientation of M32's isophote at the break (PA $\sim$ \pabreak\ deg E of N at $R_{\rm iso} \sim \risokpc$ kpc) should be pointed away from M32's direction of motion, or clockwise around M31 (PA $\sim$ \choiorbitpa\ deg E of N; C02). The sense of this inferred orbit may be naively consistent with our detection of a change in M32's rotational direction to being tentatively oriented along M32's minor axes ($\theta_{R, {\rm outer}} \sim$ \thetarouter\ deg E of N; Figure~\ref{fig:rot}) and M32's apparent structural elongation in this direction (Figure~\ref{fig:maps}). It is unclear how exactly the orbit for M32 deduced by C02 relates to those in models focused on passages of M32 through M31's disk to perturb M31's disk velocity structure and star formation history \citep{Gordon2006,Block2006,Dierickx2014}, with simulated orbits ranging from polar to nearly radial orientations. 

Modern self-consistent N-body models of the M31-M32 interaction using the most recent phase space constraints are therefore required to determine probable orbits for M32 and whether the observed kinematical signatures presented in this work (Section~\ref{sec:m32}) agree with predictions for tidal distortion and/or stripping.  For example, improved distances to M32 ($D_{\odot} \sim 772.7^{+21}_{-22}$ kpc; \citealt{Savino2022}) place the results of \citealt{Dierickx2014} well outside the original 2$\sigma$ confidence intervals for reproducing M32's phase space coordinates. Although the uncertainties on M31's bulk transverse motion  remain significant \citep{vanderMarel2008,Sohn2012,vanderMarel2012}, the impact of assuming \textit{Gaia} proper motions \citep{vanderMarel2019,Salomon2021} should be considered in new M31-M32 interaction models. In conjunction with these factors, upcoming HST-based proper motions for M32 and accompanying orbital models (Fardal et al., in preparation; Patel et al., in preparation) will provide 
a crucial piece of evidence regarding whether M32 is indeed inconsistent with sharing the same progenitor as the GSS and its associated tidal shells \citep{Fardal2013,Escala2022}.

In addition to providing kinematical constraints for more sophisticated modeling of the interaction between M32 and M31, we anticipate that the outer velocity structure for M32 presented in this work will enable updated models of M32's dynamical mass, as well as revisiting the tentative evidence for a dark matter halo in M32 (H13). The methodology for kinematically separating M32 from stellar populations in M31 detailed here (following \citealt{Gilbert2022} and \citealt{Cullinane2023}) will also be useful for disentangling M32's resolved stellar chemical abundance distribution from M31. This will shed further light on formation scenarios for M32, particularly in comparison to existing chemical abundance measurements in M31's inner stellar halo and tidal substructures \citep{Gilbert2019,Escala2019,Escala2020a,Escala2020b,Escala2021,Wojno2023}. Finally, the methodology could also be applied across M31's disk to provide a high-fidelity separation from M31's rotating inner halo (c.f.\ \citealt{Dorman2012,Escala2023}) toward exploring the disk--halo connection in M31 and implications for the merger history of the M31 system.

\begin{acknowledgments}
The authors thank Ekta Patel and Roeland van der Marel for insightful discussions on M32's orbit and mass modeling, David W.\ Hogg for useful discussions on statistical modeling, Charis Tsakonas, Magda Arnaboldi, and Fran\c{c}ois Hammer for enlightening conversations on the role of M32 in a major merger scenario, and Anil Seth for contributions to the slitmask design.
We also thank interns David Carter Oropeza, Ali Behram Gumus, Derek Maeshiro, and Adrian Torres from the Science Internship Program (SIP) at the University of California, Santa Cruz, for assisting with velocity measurements.

IE acknowledges financial support from programs HST GO-15891, GO16235, and GO-16786, provided by NASA through a grant from the Space Telescope Science Institute, which is operated by the Association of Universities for Research in Astronomy, Inc., under NASA contract NAS 5-26555, and from a Carnegie-Princeton Fellowship through Princeton University.

We are grateful to the many people who have worked to make the Keck Telescope and its instruments a reality and to operate and maintain the Keck Observatory. The authors wish to recognize and acknowledge the very significant cultural role and reverence that the summit of Maunakea has always had within the indigenous Hawaiian community.  We are most fortunate to have the opportunity to conduct observations from this mountain.
\end{acknowledgments}

%

\vspace{5mm}
\facilities{HST(ACS,WFPC2), Keck(DEIMOS), CFHT(MegaCam)}


\software{astropy \citep{astropy13,astropy18,astropy22}, DYNESTY \citep{Speagle2020,Koposov2024}, numpy \citep{numpy2011,Harris2020}, scipy \citep{scipy}, matplotlib \citep{Hunter2007}}



\appendix

\section{Alternate Kinematical Models}

\subsection{Models Assuming Uniform Priors}
\label{sec:appendix}

\begin{figure*}
    \centering
    \includegraphics[width=\textwidth]{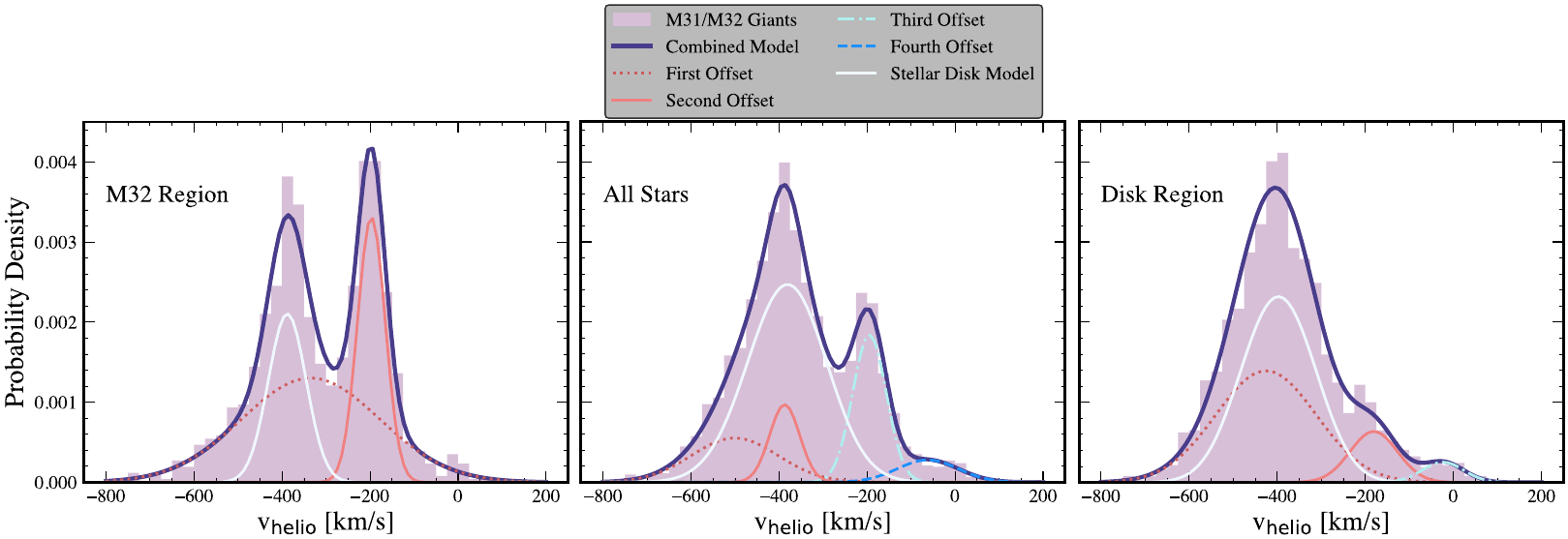}
    \caption{Same as Figure~\ref{fig:vmodel}, except assuming uniform priors agnostic of the physical nature of the ``offset'' components for the best-fit velocity models (Appendix~\ref{sec:appendix}). \added{We assume a single rotating component for M31's disk (Appendix~\ref{sec:altmods}), which encompasses the secondary disk-like component identified from all M31/M32 giant stars (middle panel; red solid line).} \added{When considering all samples,} we find evidence for a kinematically cold component corresponding to M32 and two kinematically hot components with mean velocities shifted toward the negative and positive extrema of the velocity distributions. This motivates our fiducial assumption of four total offset components when considering our prior physical knowledge of the system, where we interpret the two kinematically hot components as a M31 halo-like component plus two kinematically colder components at the positive (MW-like) and negative (GSS-like) velocity tails of the distribution (Section~\ref{sec:ncomp},~\ref{sec:prior}).
    }
    \label{fig:vmodel_unif}
\end{figure*}

Here, we provide technical details on the ``rotating plus offset'' velocity models (Section~\ref{sec:vmodel}) assuming uniform priors on all parameters discussed in Section~\ref{sec:ncomp}. In the dynamic nested sampling implementation using DYNESTY \citep{Speagle2020}, we allowed \frot\ to vary between 0 and 1.5, bounded $f_v$ between 0 and 1, \muv\ between $-800$ and $+200$ \kms, and assumed an upper limit on $\sigma_v$ of 250 \kms. We also adopted the following constraints: $\Sigma_N f_v^N = 1$, $\sigma_{v, {\rm disk}} < \sigma_{v, {\rm halo}}$, and 5 \kms\ $< \sigma_{v, N} < \sigma_{v, {\rm disk}}$, and $\mu_v^N < \mu_v^{N-1}$ to prevent degeneracy between model components in the sampling procedure. We sampled all models over \textit{identical} prior volumes, regardless of whether all parameters were used in the evaluation of a given likelihood function, to enable comparisons of the Bayesian evidence ($\mathcal{Z}$).
Table~\ref{tab:uniform} summarizes the various $N$-component models, including the change in the evidence estimation for each additional offset component, $\Delta \log \mathcal{Z} = \log \mathcal{Z}_N - \log \mathcal{Z}_{N-1}$. We increased the number of model components until $\Delta \log \mathcal{Z} $ became negative, which indicates that $N$ components is not preferred over $N-1$ components based on the Bayes factor thresholds for interpreting model evidence by \citet{KassRaftery1995} (see also Table~2 of \citealt{Cullinane2023}). For models with positive $\Delta \log \mathcal{Z}$, we considered the $N$-component model a better description of the data if there was ``positive/substantial'' evidence in favor of this hypothesis when compared to the $N-1$ component model  ($\Delta \log \mathcal{Z} \geq 0.5$). We find ``decisive'' evidence ($\Delta \log \mathcal{Z} \geq 2.2$) preferring \added{5D}, \added{4D}, and 3D models for all M31/M32 giant stars, stars in the ``Disk Region'' (\psb\ $<$ 0.01), and stars in the ``M32 Region'' (\psb\ $>$ 0.01) respectively. We discuss the physical interpretation of these model components in Section~\ref{sec:ncomp}, where Figure~\ref{fig:vmodel_unif} shows the best-fit velocity models assuming uniform priors (Table~\ref{tab:uniform}) for each spatial sample.

\begin{table*}
    \centering
    \label{tab:uniform}
    \caption{Kinematical Parameters for Rotating Plus Offset Model -- Uniform Priors}
    \begin{tabular*}{\textwidth}{@{\extracolsep{\fill}}l|cc|ccccc}
        \hline\hline
        Model &  $f_{\rm rot}$ & $\sigma_{\rm disk}$ [km s$^{-1}$] & Comp. & $f_v$ & $\mu_v$ [km s$^{-1}$] & $\sigma_v$ [km s$^{-1}$] & $\log\mathcal{Z}_{N} - \log\mathcal{Z}_{N-1}$ \\ \hline
                 \multicolumn{1}{c}{} & \multicolumn{2}{c|}{Disk} &  \multicolumn{4}{c}{Offset Components} &  \\ \hline
                 
        \multicolumn{8}{c}{All Stars}\\ \hline
        2D & $1.03^{+0.03}_{-0.27}$ & $49.5^{+64.0}_{-7.9}$ & First & $0.73^{+0.02}_{-0.62}$ & $-328.3^{+130.1}_{-7.4}$ & $146.0^{+2.2}_{-27.7}$ & +16.7 \\ \hline
        3D & $1.00^{+0.01}_{-0.03}$ & $63.2^{+3.3}_{-7.1}$ & First & $0.49^{+0.03}_{-0.06}$ & $-341.7^{+3.5}_{-7.4}$ & $160.7^{+2.5}_{-5.2}$ & +48.2 \\
        & & & Second & $0.11^{+0.01}_{-0.01}$ & $-194.1^{+1.4}_{-3.4}$ & $28.0^{+1.4}_{-3.0}$ & \\ \hline
        4D & $0.85^{+0.02}_{-0.04}$ & $74.3^{+3.3}_{-9.1}$ & First & $0.23^{+0.04}_{-0.07}$ & $-456.2^{+13.1}_{-32.0}$ & $109.5^{+4.5}_{-10.3}$ & +11.8 \\
        & & & Second & $0.18^{+0.01}_{-0.02}$ & $-193.5^{+1.5}_{-3.4}$ & $37.7^{+1.7}_{-3.8}$  \\
        & & & Third & $0.05^{+0.01}_{-0.01}$ & $-57.0^{+8.9}_{-21.3}$ & $65.3^{+3.9}_{-9.2}$ \\ \hline
        5D & $0.85^{+0.03}_{-0.05}$ & $82.7^{+3.2}_{-9.7}$ & First & $0.14^{+0.02}_{-0.03}$ & $-500.7^{+14.3}_{-25.3}$ & $98.2^{+5.7}_{-10.8}$ & +2.5 \\
        & & & Second & $0.08^{+0.02}_{-0.03}$ & $-387.3^{+4.2}_{-8.8}$ & $34.2^{+4.5}_{-9.3}$ & \\
        & & & Third & $0.18^{+0.01}_{-0.03}$ & $-194.5^{+1.6}_{-3.8}$ & $38.2^{+2.2}_{-4.8}$ & \\
        & & & Fourth & $0.05^{+0.01}_{-0.01}$ & $-62.0^{+11.4}_{-26.8}$ & $69.1^{+5.1}_{-12.2}$ & \\ \hline
        
        \multicolumn{8}{c}{Disk Region, $p_{\rm M32, SB} < 0.01$} \\ \hline 
        2D &  $0.98^{+0.02}_{-0.04}$ & $70.6^{+2.6}_{-6.4}$ & First & $0.47^{+0.02}_{-0.06}$ & $-344.8^{+4.5}_{-9.2}$ & $161.7^{+2.9}_{-6.2}$  & +56.6 \\ \hline
        3D & $0.95^{+0.03}_{-0.09}$ & $74.1^{+5.4}_{-10.4}$ & First & $0.45^{+0.05}_{-0.16}$ & $-375.9^{+8.2}_{-49.8}$ & $138.9^{+5.1}_{-22.6}$   & +1.3 \\
        & & & Second & $0.04^{+0.03}_{-0.02}$ & $-65.4^{+32.3}_{-56.7}$ & $65.5^{+9.2}_{-28.8}$  & \\ \hline
        4D & $0.90^{+0.02}_{-0.06}$ & $68.8^{+3.2}_{-7.4}$ & First & $0.40^{+0.04}_{-0.09}$ & $-425.5^{+8.2}_{-23.0}$ & $113.9^{+3.9}_{-9.0}$   & $+$3.7\\ 
        & & & Second & $0.09^{+0.01}_{-0.02}$ & $-180.9^{+5.7}_{-13.2}$ & $53.9^{+4.7}_{-11.2}$  & \\
        & & & Third & $0.03^{+0.00}_{-0.01}$ & $-23.8^{+8.0}_{-23.1}$ & $47.0^{+5.7}_{-10.9}$ & \\ \hline
        5D & $0.91^{+0.02}_{-0.06}$ & $68.6^{+3.1}_{-7.0}$ & First & $0.37^{+0.04}_{-0.11}$ & $-431.1^{+9.5}_{-31.2}$ & $112.8^{+4.0}_{-10.4}$ & $-1.3$ \\
        & & & Second & $0.06^{+0.02}_{-0.04}$ & $-212.0^{+18.9}_{-172.1}$ & $49.6^{+6.2}_{-18.9}$ & \\
        & & & Third & $0.05^{+0.03}_{-0.04}$ & $-167.5^{+30.8}_{-25.2}$ & $52.8^{+5.3}_{-15.2}$ & \\
        & & & Fourth & $0.03^{+0.00}_{-0.01}$ & $-18.6^{+8.5}_{-24.6}$ & $45.3^{+6.1}_{-12.1}$ & \\ \hline
        
        \multicolumn{8}{c}{M32 Region, $p_{\rm M32, SB} > 0.01$} \\ \hline   
        2D & $1.18^{+0.07}_{-0.12}$ & $28.2^{+5.7}_{-7.1}$ & First & $0.90^{+0.01}_{-0.35}$ & $-301.5^{+4.6}_{-6.5}$ & $137.1^{+2.0}_{-38.1}$   & $+$1.9 \\ \hline
        3D &  $1.08^{+0.03}_{-0.06}$ & $41.3^{+3.1}_{-6.1}$ & First & $0.51^{+0.03}_{-0.07}$ & $-334.9^{+4.1}_{-9.7}$ & $156.9^{+3.2}_{-6.7}$   & +75.6 \\
        & & & Second & $0.26^{+0.01}_{-0.03}$ & $-197.0^{+1.3}_{-3.0}$ & $31.5^{+1.3}_{-2.9}$ 
 & \\\hline
        4D & $1.07^{+0.03}_{-0.08}$ & $41.5^{+3.8}_{-6.5}$ & First & $0.49^{+0.03}_{-0.10}$ & $-341.7^{+5.4}_{-16.8}$ & $153.6^{+4.0}_{-9.0}$  & $-1.8$ \\
        & & & Second & $0.25^{+0.02}_{-0.07}$ & $-198.4^{+1.8}_{-7.2}$ & $31.3^{+2.0}_{-5.0}$  & \\
        & & & Third & $0.02^{+0.02}_{-0.01}$ & $-55.2^{+42.5}_{-128.9}$ & $28.2^{+4.9}_{-13.0}$  & \\
        \hline
\end{tabular*}
\end{table*}

\subsection{Models with Multiple Rotating Components}
\label{sec:altmods}

In this section, we evaluate whether models with multiple rotating components, such as a two-component disk and/or a rotating stellar halo, could provide a better description of the observed velocity distribution than the fiducial model composed of a single rotating disk component and an offset stellar halo (Section~\ref{sec:vmodel}). 
\added{The statistical evidence in favor of a potential secondary disk component (Section~\ref{sec:ncomp}; Appendix~\ref{sec:appendix})}, as well as the apparent asymmetry near the disk velocity peak extending toward more negative velocities (Figure~\ref{fig:vmodel}), provides reason to explore the possibility of a two-component disk structure. Moreover, \citet{Dorman2012} found signatures of rotation in M31's stellar halo, with $V_{\rm rot,halo} = 52.6 \pm 6.8$ \kms\ for $R_{\rm proj}$ $>$ 5 kpc approximately from M31's southwest to northeast major axis (the same direction as M31's disk rotation). Assuming that M31's halo rotates in the plane of the disk with $V_{\rm rot, halo, max}$ = 52.6 \kms, its predicted median line-of-sight velocity is $v_{\rm los, halo}$ = \vloshalorot\ \kms\ over the DEIMOS footprint (from an analogous modification of Equation~\ref{eq:vmod}), where $v_{\rm los, halo}$ has predicted maximum and minimum values of \vloshalorotmax\ and \vloshalorotmin\ \kms\ over this area.

To implement multiple rotating components, we modified the likelihood function in Equation~\ref{eq:likelihood} accordingly and sampled the posterior distribution using DYNESTY \citep{Speagle2020}. For a two-component disk structure, we assumed kinematically ``thinner'' and ``thicker'' disks, following the constraints $\sigma_{v, {\rm thin}} < \sigma_{v, {\rm thick}}$ and $f_{\rm rot, thin} > f_{\rm rot, thick}$. Given that the MW-like and GSS-like model components combined contribute $\sim$5\% of the total stellar population (Table~\ref{tab:offset_model_norm}), we omitted these features in this modeling exercise for simplicity, considering models composed only of M32, M31's stellar halo, and M31's stellar disk(s). We enforced normal priors on M32 and M31's halo velocity dispersion (Table~\ref{tab:prior}), but lifted the normal priors on M31's disk velocity dispersion, instead requiring that $\sigma_{v, {\rm thick}} < \sigma_{v, {\rm halo}}$ in addition to the aforementioned constraints on the disk(s). In the case of a non-rotating ``offset'' halo, we imposed normal priors on $\mu_{v, {\rm halo}}$ (Table~\ref{tab:prior}). For a rotating halo, we adopted the constraints $f_{\rm rot, halo} < f_{\rm rot,thick}$ in the case of a two-component disk structure and $f_{\rm rot, halo} < f_{\rm rot,disk}$ otherwise. We also implemented the Gaussian prior $\mathcal{N}(0.2|0.05)$ on $f_{\rm rot,halo}$ based on the predicted values of $V_{\rm rot, halo}$/$V_{\rm HI, rot}$ over the DEIMOS footprint, and retained the prior on $f_{\rm halo}$ (Table~\ref{tab:prior}) for both rotating and non-rotating halo models. 
We estimated the Bayesian evidence $\mathcal{Z}$ over this fixed prior volume for \textit{all} model parameters (i.e., for a two-component disk, rotating halo, and non-rotating halo) and resampled the posterior distribution of the standard 3D model (Table~\ref{tab:offset_model_norm}) under these prior assumptions.

We found that, under the assumption of an offset halo, a two-component disk is decisively preferred ($\Delta \log \mathcal{Z} \sim$ \dlogztwovsonediskoffsethalo). 
\added{When assuming a second rotating component, a model with with a two-component disk and an offset halo is decisively preferred over a model with a single-component disk and rotating halo ($\Delta \log \mathcal{Z} \sim$ \dlogztwodiskoffsetvsonediskrot).}
Under the assumption of a rotating halo, a two-component disk model is also decisively preferred ($\Delta \log \mathcal{Z} \sim$ \dlogztwodiskrotvsoffsethalo). In this model, $\sigma_{v, {\rm thin}} \sim$ \sigmathin\ \kms\ and $\sigma_{v, {\rm thick}} \sim$ \sigmathick\ \kms\ with $f_{\rm rot,thin} \sim$ \frotthin\ and $f_{\rm rot,thick} \sim$ \frotthick\ for the ``thin'' and ``thick'' disk components, where $f_{\rm rot, halo} \sim$ \frothalo\ and $\sigma_{v, {\rm halo}} \sim$ \sigmahalorot\ \kms\ for the rotating halo. Thus, the addition of a ``thin'' disk does not reduce the velocity dispersion of the ``thick'' disk, nor does it account for the negative tail of the observed velocity distribution. \added{Instead, the model parameters corresponding to the ``thin'' and ``thick'' disks are qualitatively similar to those for models fit separately to the AGB and RGB populations (Table~\ref{tab:rgb}; Appendix~\ref{sec:vmodel_rgb}).}

Given that M31's disk and halo components effectively provide nuisance parameters over which we marginalize to extract M32's line-of-sight velocity distribution, we assumed a model composed of a single-component disk and offset halo in this work for simplicity. The computed M32 probabilities for a given star (Equation~\ref{eq:likelihood}) are strongly correlated under each of the four models explored here, and therefore do not significantly alter our main results regarding M32 (Section~\ref{sec:m32}). A more detailed analysis over a larger and more representative area of M31's southern disk region, free of the added complexity of M32, is required to confirm whether its velocity distribution supports a rotating halo or a multiple-component disk structure and is therefore beyond the scope of this paper.


\subsection{Models with Stellar Age}
\label{sec:vmodel_rgb}

\begin{figure}
    \centering
\includegraphics[width=\textwidth]{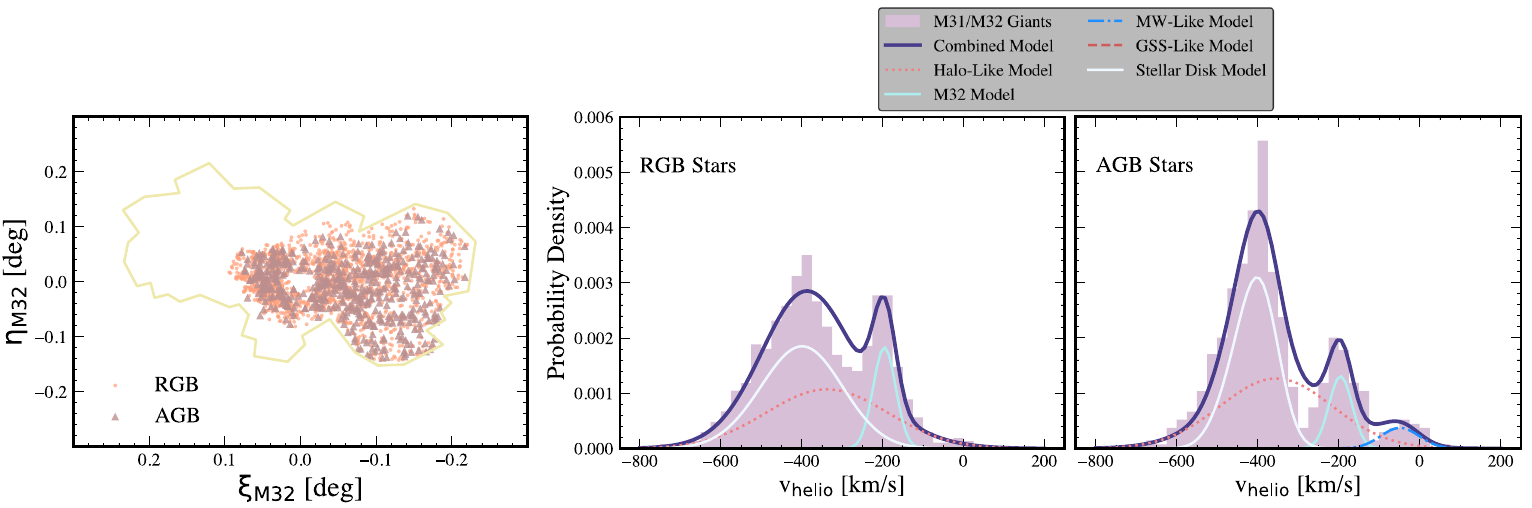}
    \caption{(Left) M32-centric sky location of RGB (orange points) and AGB (brown triangles; Section~\ref{sec:rgb}) stars, whose classification is limited to the PHAST footprint (compared to the DEIMOS footprint in Figure~\ref{fig:obs}; yellow outline). (Right panels) Similar to Figure~\ref{fig:vmodel}, except for the RGB and AGB star velocity distributions (Appendix~\ref{sec:vmodel_rgb}). M32 is primarily traced by the old RGB population, \added{although} a statistically significant M32 contribution is \added{also} 
    detected in the AGB stars. 
    More AGB than RGB stars are present at MW-like velocities. Neither the RGB nor AGB stars favor a significant GSS-like component despite the 
    \added{presence} of stars at $-600$ \kms. 
    \added{The primary difference between the RGB and AGB models is the velocity dispersion of M31's disk, where the hotter kinematics of the RGB stars encapsulates the AGB stars. The RGB model is broadly consistent with a version of the 5D model restricted to the PHAST footprint (Table~\ref{tab:rgb}), excepting the marginally higher velocity dispersion of the RGB model, which may be artificially inflated due to the absence of the GSS-like population.}
    }
    \label{fig:vmodel_rgb}
\end{figure}

\begin{table}
    \centering
    \caption{Kinematical Parameters for Rotating Plus Offset Models by Stellar Type -- Normal Priors}
    \label{tab:rgb}
    \begin{tabular*}{\textwidth}{@{\extracolsep{\fill}}l|cc|ccccc}
        \hline\hline
        Model &  $f_{\rm rot}$ & $\sigma_{\rm disk}$ [km s$^{-1}$] & Comp. & $f_v$ & $\mu_v$ [km s$^{-1}$] & $\sigma_v$ [km s$^{-1}$] \\ \hline
         \multicolumn{1}{c}{} & \multicolumn{2}{c|}{Disk} &  \multicolumn{3}{c}{Offset Components} &  \\ \hline

        \multicolumn{7}{c}{PHAST Survey Area (RGB \& AGB Stars)}\\ \hline
        5D & $0.97^{+0.02}_{-0.04}$ & $74.7^{+3.0}_{-6.6}$ & Halo-Like & $0.36^{+0.02}_{-0.05}$ & $-347.9^{+4.6}_{-10.0}$ & $142.6^{+2.7}_{-6.1}$ \\
        & & & M32 & $0.14^{+0.01}_{-0.01}$ & $-194.7^{+1.1}_{-2.6}$ & $29.6^{+1.1}_{-2.4}$ \\
        & & & MW-Like & $0.02^{+0.00}_{-0.01}$ & $-52.6^{+3.9}_{-9.5}$ & $54.2^{+3.1}_{-8.0}$ \\
        & & & GSS-Like & $0.01^{+0.00}_{-0.01}$ & $-610.0^{+4.3}_{-11.6}$ & $34.3^{+3.5}_{-8.1}$ \\ \hline
         
        \multicolumn{7}{c}{RGB Stars}\\ \hline
        3D & $0.95^{+0.02}_{-0.06}$ & $94.3^{+2.9}_{-6.8}$ & Halo-Like & $0.40^{+0.02}_{-0.05}$ & $-337.5^{+3.4}_{-7.7}$ & $150.2^{+2.2}_{-5.0}$ \\
        & & & M32 & $0.13^{+0.01}_{-0.02}$ & $-195.3^{+1.3}_{-2.9}$ & $27.8^{+1.3}_{-2.8}$  \\
        
        \hline
        \multicolumn{7}{c}{AGB Stars}\\ \hline
        4D & $0.99^{+0.02}_{-0.03}$ & $41.1^{+3.8}_{-6.5}$ &  Halo-Like & $0.44^{+0.02}_{-0.05}$ & $-354.4^{+4.2}_{-9.5}$ & $138.2^{+3.2}_{-8.0}$  \\
        & & & M32 & $0.09^{+0.01}_{-0.02}$ & $-195.5^{+1.7}_{-3.9}$ & $28.5^{+1.6}_{-3.6}$  \\
        & & & MW-Like & $0.04^{+0.01}_{-0.01}$ & $-48.7^{+3.9}_{-8.9}$ & $47.2^{+3.5}_{-8.1}$ \\
         \hline
\end{tabular*}
\end{table}

In Section~\ref{sec:kinematics}, we modeled the kinematics of the entire sample of M31/M32 giant stars (regardless of stellar classification) in order to maximize our sample size and spatial coverage, assuming that the hotter kinematics of the older RGB population encompasses that of the kinematically colder and younger AGB stars. To explore kinematical differences with respect to stellar age, we separated our spectroscopic sample into RGB and AGB populations (Section~\ref{sec:rgb}), then repeated the kinematical modeling procedure for the RGB and AGB samples. 
We also modeled the combined RGB and AGB sample (i.e., all stars with stellar classifications as opposed to the full sample of all M31/M32 giant stars) to control for the areal coverage of this subsample, which traces the PHAST survey area. 
\added{We found results mostly consistent with Table~\ref{tab:offset_model_norm}, excepting marginal evidence for a larger M31 disk dispersion (1.8$\sigma$), lower M31 halo fraction ($1.3\sigma$), and higher M32 fraction (1.1$\sigma$) over the PHAST versus DEIMOS footprints. This suggests that any kinematical differences between the RGB and AGB stars beyond those associated with the few aforementioned model parameters are not driven by survey area.
}
We show the RGB and AGB velocity distributions, as well as their spatial distributions, 
in Figure~\ref{fig:vmodel_rgb}. The RGB velocities show clear peaks associated with M31's stellar disk and M32, and a negatively skewed asymmetry in the disk velocity peak that may suggest the presence of stars at GSS-like velocities or possibly multiple disk components (Appendix~\ref{sec:altmods}). The AGB velocity distribution appears dominated by M31's stellar disk, with small peaks at $-600$ and $-200$ \kms\ and a more prominent tail toward MW-like velocities.

For the RGB stars, we found evidence for M31's stellar disk and \added{three} offset model components corresponding to M32, a kinematically hot component with halo-like dispersion (\sigmav\ $\sim$ \added{115} \kms) but mean velocity shifted toward the \added{negative} tail of the distribution (\muv\ $\sim$ $-$\added{470} \kms), \added{and a kinematically hot component at the positive velocity tail (\muv\ $\sim$ $-100$ \kms, \sigmav\ $\sim$ 85 \kms)}. We thus initially 
assumed an RGB model with \added{five} total components: M31's stellar disk, an M31 halo-like component, M32, a MW-like component, \added{and a GSS-like population}. For the AGB stars, we found evidence for M31's stellar disk and 
\added{three offset components with similar characteristics as those identified from the RGB stars, with the exception of the component at the positive velocity tail having a less negative mean velocity and lower dispersion (\muv\ $\sim-30$ \kms, \sigmav\ $\sim$ 40 \kms).}
We thus assumed a \added{five} component AGB model consisting of M31's stellar disk, an M31-halo like population, \added{M32}, a MW-like population, \added{and a GSS-like population.} 

When assuming the normal priors from Table~\ref{tab:prior}, including the constraints on the fractional halo-like contribution $f_h$, the best-fit RGB model supports three components: M31's stellar disk, M32, and an M31 halo-like component. This differs from the \added{five} components inferred from uniform priors, likely because the comparatively small number of RGB stars at 
tails of the velocity distribution are subsumed into the high-dispersion halo.
We confirmed that including a MW-like component ($\Delta \log \mathcal{Z} \sim$ \dlogzrgbmw) or GSS-like component ($\Delta \log \mathcal{Z} \sim$ \dlogzrgbgss) is not preferred for the RGB velocity distribution. 
In the case of normal priors, the AGB velocity distribution is 
best described by \added{a 4D} model including the MW-like component, as opposed to a simpler model consisting solely of M31's stellar disk and halo ($\Delta \log \mathcal{Z} \sim$ \dlogzagbfourdvstwod) \added{or additionally including M32 but not the MW- or GSS-like components ($\Delta \log \mathcal{Z} \sim$ \dlogzagbdwarf)}. 
We verified that \added{including a GSS-like component is disfavored ($\Delta \log \mathcal{Z} \sim$ \dlogzagbgss) compared to the standard 4D AGB model, again suggesting that the small number of stars with GSS-like velocities are subsumed into a separate kinematical component.}

Table~\ref{tab:rgb} summarizes the kinematical parameters for the adopted RGB and AGB models (Figure~\ref{fig:vmodel_rgb}). The RGB and AGB stars rotate with similar speeds as the HI gas, where the RGB stars slightly lag the HI disk (\frot\ $\sim -0.95$) and have a \added{substantially} larger dispersion (\sigmav\ $\sim$ \added{90} \kms) compared to the AGB stars. \added{The M32 fraction is higher in the RGB population than the AGB population, suggesting that our spectroscopic sample predominately traces the old population in M32 (Section~\ref{sec:rgb}).}
Moreover, the dispersion of the M31 halo-like component may be higher for the RGB stars than the AGB stars. 
\added{The main difference between the RGB and AGB velocity distributions is therefore the M31 disk dispersion, where our initial assumption that the hotter kinematics of the RGB stars encapsulates the AGB stars is valid over our survey area}, supporting our choice to model the entire M31/M32 giant distribution in Section~\ref{sec:kinematics}.

Compared to the 5D model for all M31/M32 giant stars \added{over the PHAST survey area (Table~\ref{tab:rgb})}, M31's disk dispersion is marginally higher for 
the RGB 
model by $\sim$\sigmargbsigmadisk$\sigma$. 
This may be driven by an intrinsically higher disk dispersion compared to a sample containing AGB stars and/or the putative GSS-like population being subsumed into the rotating model component in the absence of a dedicated GSS-like component in the adopted 
RGB model, thereby inflating the modeled disk dispersion. 
\added{The AGB model has a lower M31 disk dispersion, M31 halo dispersion, and M32 fraction compared to the 5D PHAST model (as expected from the earlier comparison to the RGB model), with the only other notable difference in the AGB model being a marginally higher contribution of MW-like stars (1.8$\sigma$).}

We defer further modeling and analysis of kinematical differences between RGB and AGB populations in M32, M31's stellar disk, M31's stellar halo, and possible M31 substructure to future work, ideally based on spectroscopic samples with more comprehensive spatial coverage of M31's southern disk region (Grion Filho et al., in preparation) and with larger samples of giant stars with available PHAST photometry.

\section{Predictions for GSS Merger-Related Tidal Debris}
\label{sec:gss}

\begin{figure}
    \centering
    \includegraphics[width=0.5\columnwidth]{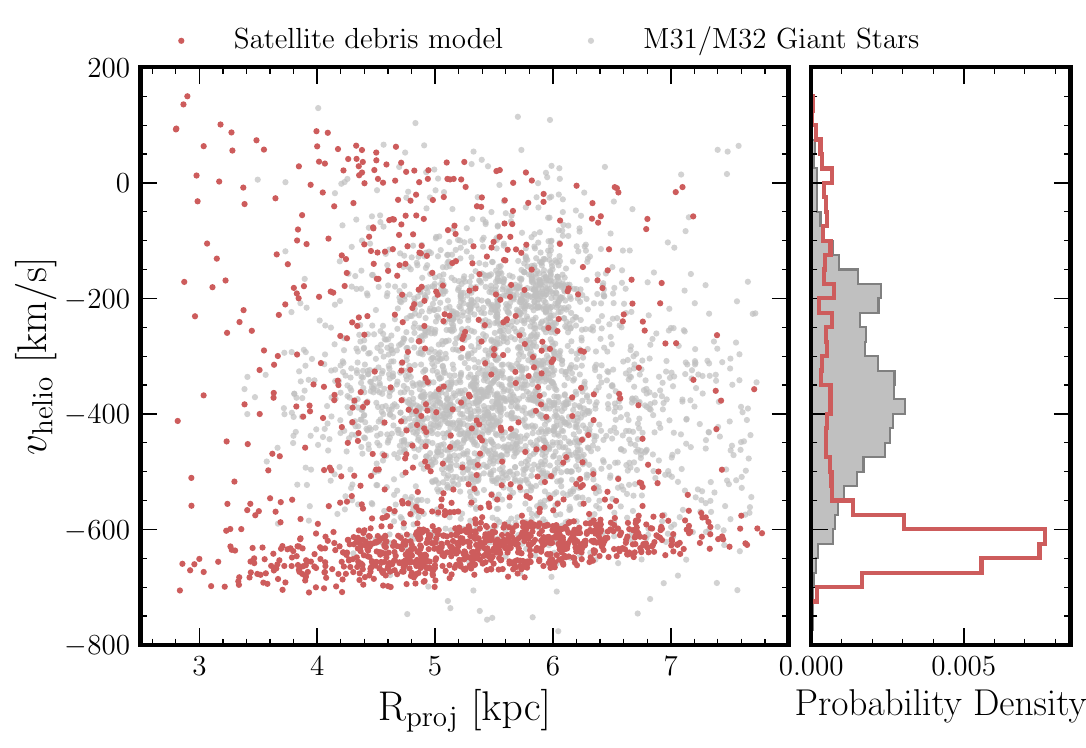}
    \caption{Heliocentric velocity versus projected M31-centric radius over the DEIMOS footprint (Figure~\ref{fig:obs}) for M31/M32 giant stars (silver points; Section~\ref{sec:m31bayes}) and satellite debris particles from an N-body model for the formation of the GSS in a minor merger (red points;  \citealt{Fardal2012,Escala2022}; Section~\ref{sec:prior}). Observational velocity errors are not shown for clarity (median \vtotmed\ \kms). The model predicts M31 to dominate the stars in this region, though GSS-related tidal debris should be most detectable near $-600$ \kms\ (Table~\ref{tab:prior}). The predicted debris also extends to MW-like velocities (near 0 \kms).
    }
    \label{fig:nbody}
\end{figure}

Here, we present Figure~\ref{fig:nbody} demonstrating predictions for satellite tidal debris from the GSS progenitor in a minor merger scenario \citep{Fardal2012,Escala2022}, which we use to inform our prior selection in Section~\ref{sec:prior}.

\end{document}